\documentclass[12pt]{article}
\usepackage{amssymb}
\usepackage{amsmath}
\usepackage{hyperref}
\numberwithin{equation}{section}
\begin{document}
\title{\bf Entanglement entropy of singular surfaces under relevant deformations in holography}
\author{Mostafa Ghasemi\thanks{Email: m.ghasemi.g@modares.ac.ir}  \hspace{2mm} and
        Shahrokh Parvizi\thanks{Corresponding author: Email: parvizi@modares.ac.ir}\hspace{1mm}\\
		{\small {\em  Department of Physics, School of Sciences,}}\\
        {\small {\em Tarbiat Modares University, P.O.Box 14155-4838, Tehran, Iran}}\\
       }
\date{\today}
\maketitle

\abstract{In the vacuum state of a $CFT$, the entanglement entropy of singular surfaces contains a logarithmic universal term which is only due to the singularity of the entangling surface. We consider the relevant perturbation of a three dimensional $CFT$ for singular entangling surface. We observe that in addition to the universal term due to the entangling surface, there is a new logarithmic term which corresponds to a relevant perturbation of the conformal field theory with a coefficient depending on the scaling dimension of the relevant operator. We also find a new power law divergence in the holographic entanglement entropy. In addition, we study the effect of a relevant perturbation in the  Gauss-Bonnet gravity for a singular entangling surface. Again a logarithmic term shows up. This new term is proportional to both the dimension of the relevant operator and the Gauss-Bonnet coupling. We also introduce the renormalized entanglement entropy for a kink region which in the $UV$ limit reduces to a universal positive finite term.

\vspace{1cm}
%\pacs{11.10.Jj, 11.10.Wx, 11.15.Pg, 11.25.Tq}


%--------------------------------------------------------------------------
\section{Introduction} \label{intro}
Entanglement entropy is one of the important measures of entanglement feature in quantum systems which emerges in diverse research area \cite{Ref1,Ref2,Ref3,Ref4,Ref5,Ref6,Ref7,Ref8,Ref9,Ref10,Ref11,Ref12,Ref13}. In the context of quantum field theory, the entanglement entropy of a sub region ${A}$ is defined as Von Neumann entropy of reduced density matrix $\rho_{A}$, i.e., $ S=-Tr( \rho_{A} log\rho_{A})$. The reduced density matrix is in turn defined by tracing over the degrees of freedom of complementary region $\bar{A}$ of $ {A}$, $\rho_{A}=Tr_{\bar{A}}  (\rho)$. In general, the entanglement entropy suffers from the $UV$ divergence due to the short range correlation across the so-called entangling surface, the boundary of two regions. So in order to have a well-defined quantity it must be regularized.

In the vacuum of a $(2+1)$- dimensional $CFT$, the entanglement entropy for a smooth entangling surface takes the following form

\begin{equation}%\label[1]{1}
  % S_EE=α l_Σ/δ-F,                                                                       (1.1)
    S_{EE}=a \frac{l_\Sigma}{\delta} -F,
\end{equation}
where $l_\Sigma$ is the length of the entangling surface, $\delta$ is a $UV$ cut-off, and $a$ is a constant depending on the details of the underlying theory. The leading term exhibits the “area law” \cite{Ref14,Ref15} and the second term $F$ is a universal term independent of the regularization scheme. In the special case, when entangling surface is a circle of radius $R$, $F$ is related to the renormalized Euclidean partition function $Z$ of the $CFT$ on the three dimensional sphere $S^3$, $F=-log|Z|$ \cite{Ref16}. The $F$ term actually appears in a constraint by an $F$-theorem \cite{Ref17,Ref18}, which states that in a unitary quantum field theory, $F$ is positive, stationary at fixed points, and monotonically decreases along an RG flow between $UV$ and $IR$ fixed points, $F_{UV}>F_{IR}$. In the holography context it was addressed in \cite{Ref19}. The proof of its positivity and monotonicity relies on the renormalized entanglement entropy \cite{Ref20,Ref21}, that is defined as
 \begin{equation}
  \mathcal{F}(R)=R\partial_{R}S(R)-S(R),
  \end{equation}
where $R$ is the length scale of the entangling surface. We will come back to this point in section \ref{sec5}.

\indent
In 3 dimensional $CFT$, when there is a singularity in the entangling surface, the entanglement entropy contains an additional singular term which is universal  %\cite{Ref10}.\\
%S_EE=B H/δ-a(Ω)log(H/δ)+O(1),                                      (1.2)
\begin{equation}\label{SEE-aOmega}
  S_{EE}=\beta\frac{H}{\delta}- a(\Omega)\log(\frac{H}{\delta})+O(1),
\end{equation}
where $\Omega$ is the opening angle, $\beta$ is a non-universal constant, and $H$ denotes the size of the entangling surface. $a(\Omega)$ is a coefficient of the new logarithmic term that appears due to the corner shape of the entangling surface and gives the universal part of the $EE$. It is a positive convex function that satisfies some properties (for details see \cite{Ref22,Ref23,Ref24,Ref25,Ref26,Ref27,Ref28}).

We note that similar universal logarithmic terms appear in conformal field theories in other dimensions, logarithmic terms in even dimensions \cite{Ref11,Ref29,Ref30,Ref31}, and double logarithmic terms in singular entangling surfaces in higher dimensions \cite{Ref32,Ref33}.

On the other hand, there are several studies on the relevant perturbation of conformal field theories \cite{Ref20,Ref34,Ref35,Ref36,Ref37,Ref38,Ref39,Ref40}. It is known that a universal logarithmic term shows up in the entanglement entropy when a $CFT$ is perturbed by a relevant operator. From the holographic point of view, the relevant perturbation corresponds to including a massive scalar field in the bulk which can deform the background from a pure $AdS$ space to an asymptotically $AdS$. The deviation depends on the scaling dimension of the relevant operator. It has been shown in \cite{Ref35} that in the first order of perturbation only for scaling dimension $\Delta=(d+2)/2$ the $S_{EE}$ receives a universal logarithmic term proportional to the scaling parameter. The same result can be found by the field theoretic calculations in\cite{Ref34,Ref36,Ref37,Ref38}. In this work, we raise the question that how the effects of corner singularity and the relevant perturbation may get mixed up. In the frame work of holography, we consider a $3$ dimensional $CFT$ on the boundary of an asymptotically $AdS$ space which is perturbed by a massive scalar field. We take a singular entangling surface and derive the entanglement entropy by the Ryu-Takayanagi prescription \cite{Ref10,Ref11}. The results show that we recover two independent universal logarithmic terms: One for the corner contribution to $S_{EE}$ and the other due to the relevant perturbation exactly for scaling dimension $\Delta=(d+2)/2$.

In the holographic frame work, it is important to study the higher curvature gravities where, in contrast to the Einstein gravity, usually new features appear. A minimal higher curvature modification would include addition of the Gauss-Bonnet ($GB$) term. In $4$ dimensions, this term is topological and does not contribute to the equation of motion. We therefore expect to have the same background. However, we know that the Ryu-Takayanagi prescription is modified in the presence of the $GB$ term. In the presence of a singular entangling surface, it has been shown in \cite{Ref27} that there is no contribution of $GB$ modification to the universal term in $S_{EE}$. We investigate this case in the presence of the relevant perturbation of the $CFT$ and show that indeed there is a universal term proportional to the $GB$ coupling and the scaling parameter exactly at $\Delta=(d+2)/2$.

This paper is organized as follows. In section 2, we review the holographic view to the relevant perturbation of a $CFT$. In section 3, first we briefly review the holographic entanglement entropy of a kink in an $AdS$ background, then we derive the Ryu-Takayanagi minimal surface in the perturbed asymptotic $AdS$ background. In section 4, we add the Gauss-Bonnet term and find out the entanglement entropy by plugging the minimal surface into the modified Ryu-Takayanagi formula. In section 5, we introduce renormalized version of the entanglement entropy. Results are discussed in section 6.

%-------------------------------------------------------------------------
\section{Relevant Perturbations in Holographic Framework}
\label{sec2}

\indent In this section we review the relevant perturbation of a conformal field theory in three dimensions in the $AdS/CFT$ context \cite{Ref35,Ref41,Ref42}. We consider the $CFT$ perturbed by a relevant operator $O(x)$ of scaling dimension $\Delta$,
\begin{equation}%\label{}
  I_{CFT}\rightarrow I=I_{CFT}+\lambda\int d^{3}xO(x),
\end{equation}
where $\lambda$ is the coupling constant of the relevant operator $O(x)$ and $\Delta<3$. By the holographic prescription, this relevant perturbation is described by turning on the scalar degrees of freedom in the bulk in which we have Einstein gravity coupled to a massive scalar field,
\begin{equation}%\label{}
I=\frac{1}{16\pi G_{N}}\int d^{4} x\sqrt{G}[R-\frac{6}{L^{2}}]-\frac{1}{2}\int d^{4} x\sqrt{G}[(\partial\Phi)^{2}+M^{2}\Phi^{2}].
\end{equation}

The relevant operator has a weak effect on the $UV$ regime while it strongly affects the $IR$ regime. In the holographic context, then the bulk geometry is asymptotically $AdS$ in the presence of the relevant operator. The metric of an asymptotically  $AdS_{4} $ can be taken as
\begin{equation}\label{AAdS}
 ds ^{2}= \frac{L^{2}}{z^{2}}(-dt^{2}+d\rho^{2}+\rho^{2}d\theta^{2}+\frac{dz^{2}}{f(z)}),
\end{equation}
where $f(z)\rightarrow1$ as $z\rightarrow0$ and thus encodes the deviation from $AdS$ spacetime. By the $AdS/CFT$ prescription, the mass $M$ of the scalar field and the scaling dimension $\Delta$ of the relevant operator are related by
\begin{equation}%\label{}
 \Delta _{\pm}=\frac{3}{2}\pm\sqrt{\frac{9}{4}+ML^{2}}.
\end{equation}
Near the boundary, the asymptotic expansion of the scalar field takes the form
\begin{equation}\label{Delta}
 \Phi(z,x)\rightarrow z^{\Delta_{-}}[\Phi^{(0)}+\cdots]+z^{\Delta_{+}}[\Phi^{(\Delta - \frac{3}{2})}+\cdots]
\end{equation}
The interpretation of $\Delta_{\pm}$ depends on the range of the scalar field mass or equivalently the conformal dimensions of the relevant operator. In the standard quantization, the range of scaling dimension of relevant operator is $3/2< \Delta <3$, and we choose $\Delta$ to be $\Delta_{+}$. In this range, we regard $\Phi^{(0)}$ and $\Phi^{(\Delta-3/2)}$, respectively, as the source of the coupling $\lambda$ and vacuum expectation value of the operator $\langle O \rangle$. On the other hand, for the alternative quantization of $1/2< \Delta<3/2 $ we have $\Delta=\Delta_{-}$, and in this range $\Phi^{(\Delta-3/2)}$ is identified with $\lambda$ while $\Phi^{(0)}$ with $\langle O \rangle$. When $\Delta = 3/2$, we should replace $ z^{\Delta_{-}}$ with $ z^{\Delta_{-}} \log z$  in \eqref{Delta}.

In order to take into account the back-reaction of the scalar field into the background geometry and finding the $f(x)$, we must solve the Einstein equation and the scalar wave equation. At all, for a source deformation and near the boundary, $f(z)$ can be expanded as
\begin{equation}\label{fz}
  f(z)= 1+\mu^{2\alpha}z^{2\alpha}+\cdots, \qquad z \rightarrow 0
\end{equation}
where $\mu$ is some mass scale\footnote{Since the only dimensionful parameter is the coupling $\lambda$ of the relevant operator, the dimensional analysis yields  $\mu\sim\lambda^{1/(d-\Delta)}$ \cite{Ref35,Ref39}. }, and $\alpha$ is a positive constant where for a source deformation, we have $\alpha= 3-\Delta_{+}$ in the standard quantization $3/2<\Delta<3$, and $ \alpha=\Delta_{+}$ in the alternative quantization $1/2< \Delta <3/2$. In the case for which $\Delta=3/2$, we should replace $\mu^{2\alpha}z^{2\alpha}$ in \eqref{fz} by $ (\mu z)^{3}(\log(\mu z))^{2}$.
We will consider the effect of these terms in $EE$ of a kink region.

%----------------------------------------------------------------------------------
\section{Holographic entanglement entropy of kink}
\label{sec3}

In this section, we study the entanglement entropy of a relevant perturbed conformal field theory in three dimensions where the entangling surface has a kink singularity. First we review the entanglement entropy of the kink for holographic $CFT$s dual to Einstein gravity which was derived in \cite{Ref25}. Then, we focus on the relevant perturbation of those $CFT$s.

% % % % % % % % % % % % % % % % % % % % % %
\subsection{The Entanglement in Pure $AdS$}
\label{sec31}

The kink region in the time slice $t_{E}= 0 $, is parameterized by $V=\{0\leq \rho\leq H,-\Omega/2\leq \theta \leq \Omega/2\}$, where $H$ is an $IR$ cut-off.

According to Ryu-Takayanagi ($RT$) prescription, the holographic entanglement entropy of a sub-region $V$ on the boundary theory is given by
%S_EE=Area(γ)/(4G_N ),                                                                      (3.1)
\begin{equation}%\label{}
  S_{EE}= \frac{Area(m)}{4G_{N}},
\end{equation}
where $m$ is the bulk minimal surface which is homologous to $V$ and $\partial_{m}$ matches the entangling surface $\partial_{V}$ on the boundary. The above formula holds in the case that the bulk physics is described by the Einstein gravity.

 We parametrize the  bulk minimal surface as $\rho=\rho(z,\theta)$. Hence the induced metric on the bulk minimal surface becomes
 \begin{equation}%\label{}
   ds^{2}=\frac{L^{2}}{z^{2}}\Big((1+\rho^{'2}) dz^{2}+2\rho'\dot{\rho}dzd\theta+(\dot{\rho}^{2}+\rho^{2})d\theta^{2}\Big),
 \end{equation}
 where $L$ is the $AdS$ curvature scale, $\dot{\rho}= \partial_{\theta}\rho$, and $\rho'=\partial_{z}\rho$. The holographic entanglement entropy is given by

 \begin{equation}%\label{}
   S_{EE}=\frac{1}{4G_{N}}\int dzd\theta\sqrt{\gamma}=\frac{L^{2}}{4G_{N}}\int_{\delta}^{z_{m}}dz\int_{\frac{-\Omega}{2}+\epsilon}^{\frac{\Omega}{2}-\epsilon}
   d\theta\frac{1}{z^{2}}\sqrt{(\dot{\rho}^{2}+1)\rho^{2}+\dot{\rho}^{2}},
 \end{equation}
 in which $\delta$ is a $UV$ cut-off, $\epsilon$ is an angular cut-off, and $z_{m}$ is defined such that $\rho(z_{m},0)=H$.

  Due to the scaling symmetry of the $AdS$ space in the absence of other scale in the problem, we can choose the following ansatz to parametrize the bulk minimal surface,
  \begin{equation}%\label{}
   % ρ= z/h(θ) ,                                                                      (3.3)
   \rho=\frac{z}{h(\theta)},
  \end{equation}
  where $h(\theta)$ is defined such that $h(\theta)\rightarrow0$  as $\theta\rightarrow\pm\Omega/2$.  With this ansatz, the holographic entanglement entropy becomes

 \begin{equation}\label{entropy-functional}
   S_{EE}=\frac{L^{2}}{2G_{N}}\int_{\delta}^{z_{m}}\frac{dz}{z}\int_{0}^{\frac{\Omega}{2}-\epsilon}d\theta
   \frac{\sqrt{1+h^{2}+\dot{h}^{2}}}{h^{2}},
 \end{equation}
in which $\dot{h}=\partial_{\theta}h$. By extremizing the above action we can derive the equation of motion for $h(\theta)$ which reads

\begin{equation}%\label{}
  h(1+h^{2})\ddot{h}+2\dot{h}^{2}+(1+h^{2})(2+h^{2})=0.
\end{equation}
Since there is no explicit $\theta$ dependence in the entropy functional \eqref{entropy-functional}, there is a conserved quantity $K$ as

\begin{equation}%\label{}
 K = \frac{1+h^{2}}{h^{2}\sqrt{(1+h^{2}+\dot{h}^{2})}}= \frac{\sqrt{1+h_{0}^{2}}}{h_{0}^{2}}.
\end{equation}
where $h_{0}=h(0)$ and we used $\dot{h_{0}}\equiv\dot{h}(0)=0$. In the following we trade the integral over $\theta$ for one over $h$
   \begin{equation}\label{SEE0}
  S_{EE}=\frac{L^{2}}{2G_{N}}\int_{\delta}^{z_{m}}\frac{dz}{z}\int_{h_{0}}^{h_{c}} \frac{dh}{\dot{h}}\frac{\sqrt{1+h^{2}+\dot{h}^{2}}}{h^{2}}.
   \end{equation}

Now we analyze the divergence of the above entropy functional. Near the boundary, the integrand in the asymptotic limit behaves as
\begin{equation}%\label{}
  \frac{\sqrt{1+h^{2}+\dot{h}^{2}}}{\dot{h}h^{2}}\thicksim - \frac{1}{h^{2}} - \frac{1}{2}K^{2}h^{2}+ \cdots\,.
\end{equation}
Hence we can isolate the divergent part of the integrals in the following way
\begin{align}%\label{}
  &I= \int_{\delta}^{z_{m}}\frac{dz}{z}\int_{h_{0}}^{h_{c}} \frac{dh}{\dot{h}}\frac{\sqrt{1+h^{2}+\dot{h}^{2}}}{h^{2}}
  \nonumber\\
&=\int_{\delta}^{z_{m}}\frac{dz}{z}\int_{h_{0}}^{h_{c}}dh
  [\frac{\sqrt{1+h^{2}+\dot{h}^{2}}}{\dot{h}h^{2}}+\frac{1}{h^{2}}]+\int_{\delta}^{z_{m}}\frac{dz}{z}(\frac{1}{h_{c}}
  -\frac{1}{h_{0}})
  \nonumber \\
 &= I^{(1)}+I^{(2)},
\end{align}
in which $I^{(1)}$ and $I^{(2)}$ represent the first and second integrals, respectively. Firstly we consider  $I^{(1)}$. We differentiate it with respect to the $UV$ cut-off $\delta$ and look for various divergent terms. We find
\begin{align}%\label{}
  &\frac{dI^{(1)}}{d\delta}= \frac{-1}{\delta} \int_{h_{0}}^{h_{c}(\delta)}dh
  [\frac{\sqrt{1+h^{2}+\dot{h}^{2}}}{\dot{h}h^{2}}+\frac{1}{h^{2}}]
  \nonumber\\
  &=\frac{-1}{\delta}\int_{h_{0}}^{0}dh
  [\frac{\sqrt{1+h^{2}+\dot{h}^{2}}}{\dot{h}h^{2}}+\frac{1}{h^{2}}]-
  \frac{dh_{c}(\delta)}{d\delta}[\frac{\sqrt{1+h^{2}+\dot{h}^{2}}}{\dot{h}h^{2}}+\frac{1}{h^{2}}]_{h_{c}=h_{c}(\delta)}
  +\cdots
  \nonumber\\
 & =\frac{-1}{\delta}\int_{h_{0}}^{0}dh
  [\frac{\sqrt{1+h^{2}+\dot{h}^{2}}}{\dot{h}h^{2}}+\frac{1}{h^{2}}]+\cdots.
\end{align}
 Similarly for $I^{(2)}$,
 \begin{equation}%\label{}
 \frac{dI^{(2)}}{d\delta}=-\frac{1}{\delta h_{c}(\delta)}+\frac{1}{\delta h_{0}}=-\frac{H}{\delta^{2}}+\frac{1}{\delta h_{0}}+\cdots,
 \end{equation}
 where in the last line we used $h_{c}(\delta)=\delta/H$. So we reach to \cite{Ref24}
 \begin{equation}%\label{}
S_{EE}=\frac{L^{2}}{2G_{N}}\Big\{\frac{H}{\delta}+\Big(-\int_{h_{0}}^{0}dh
(\frac{\sqrt{1+h^{2}+\dot{h}^{2}}}{\dot{h}h^{2}}+\frac{1}{h^{2}})+\frac{1}{h_{0}}\Big)\log(\delta)+\cdots\Big\}\,.
 \end{equation}
We see that there is a logarithmic divergence due to the singularity of the entangling surface.

% % % % % % % % % % % % % % % % % % % % % % % % % % %
\subsection{Minimal Surface in the Asymptotic $AdS$}
\label{sec32}
In this subsection, we study the holographic entanglement entropy for a kink region in the asymptotic anti-de Sitter space time background which corresponds to a perturbed $CFT$. Recall the metric in \eqref{AAdS},

\begin{equation}%\label{}
 ds ^{2}= \frac{L^{2}}{z^{2}}(-dt^{2}+d\rho^{2}+\rho^{2}d\theta^{2}+\frac{dz^{2}}{f(z)}).
\end{equation}
Similar to pure $AdS$ background, we parameterize the bulk minimal surface as $\rho=\rho(z,\theta)$. So the induced metric on the entangling surface in the time slice $t=0$ becomes

\begin{equation}%\label{}
 ds ^{2}= \frac{L^{2}}{z^{2}}\Big( \big(\rho^{'2}+\frac{1}{f(z)}\big)dz^{2}+2\rho'\dot{\rho}dzd\theta+(\dot{\rho}^{2}+\rho^{2})
 d\theta^{2}\Big),
\end{equation}
then
\begin{equation}%\lable{}
\mathbf{\gamma_{ij}} = \left(
\begin{array}{ccc}
\frac{L^{2}}{z^{2}}\big(\rho^{'2}+\frac{1}{f(z)}\big) & \frac{L^{2}}{z^{2}}\rho'\dot{\rho} & \\
\frac{L^{2}}{z^{2}}\rho'\dot{\rho} &\frac{L^{2}}{z^{2}}(\dot{\rho}^{2}+\rho^{2})  \\
%\vdots & \vdots & \ddots
\end{array} \right)\,,
\end{equation}
\begin{equation}%\label{}
 \sqrt{\gamma}=\frac{L^{2}}{z^{2}}\sqrt{\big(\rho^{'2}+\frac{1}{f(z)}\big)\rho^{2}+\frac{1}{f(z)}\dot{\rho}^{2}}\,.
\end{equation}
By the $RT$ prescription, the entanglement entropy is derived as
\begin{equation}\label{SEE}
S_{EE}=\frac{1}{4G_{N}}\int dzd\theta \sqrt{\gamma} = \frac{L^{2}}{4G_{N}}\int dzd\theta \frac{1}{z^{2}\sqrt{f}}\sqrt{(f\rho^{'2}+1)\rho^{2}+\dot{\rho}^{2}}\,.
\end{equation}
By extremizing the above action we derive the equation of motion for $\rho(z,\theta)$ to be
\begin{align}\label{EoM-rho}
&2fz\rho(\rho^{2}+\dot{\rho}^{2})\rho''
+2z\rho(1+f\rho^{'2})\ddot{\rho}
-4fz\rho \rho' \dot{\rho}\dot{\rho}'
+z\rho\rho'(\rho^{2}+\dot{\rho}^{2}) f'
\nonumber\\
&-2z\Big((1+f\rho^{'2})\rho^{2}+2\dot{\rho}^{2}\Big)
-4f\rho\rho'\Big((1+f\rho^{'2})\rho^{2}+\dot{\rho}^{2}\Big)=0
\end{align}
where $\ddot{\rho}=\partial_{\theta}^{2}\rho$, $\rho''=\partial_{z}^{2}\rho$, and $\dot{\rho}'=\partial_{\theta}\partial_{z}\rho$. Now consider the following first order perturbation
\begin{equation}\label{delta-rho}
  \rho(z,\theta)=\rho_{0}+\delta\rho=\frac{z}{h(\theta)}+\delta f g(\theta)z,\qquad  f=1+\delta f \,.
\end{equation}
Plugging it in \eqref{EoM-rho}, we derive the equations of motion for $h(\theta)$ and $g(\theta)$
\begin{equation}\label{EoM-h}
h(1+h^{2})\ddot{h}+2\dot{h}^{2}+(1+h^{2})(2+h^{2})=0,
\end{equation}

\begin{align}\label{EoM-g}
&2h^{3}(1+h^{2})\delta f \ddot{g}+4h^{2}\dot{h}\Big(z\delta f'+2(2+h^{2})\delta f \Big)\dot{g}
\nonumber\\
&+2h\Big[(\delta f''z^{2}+2z\delta f')(h^{2}+\dot{h}^{2})+(2z\delta f'+(3+h^{2})\delta f)(-\ddot{h}h+2\dot{h}^{2})
\nonumber\\
&-2(z\delta f'+2\delta f)(2\dot{h}^{2}+2h^{2}+3)-2h^{2}(2+h^{2})\delta f \Big]g
\nonumber\\
&+\Big[(h^{2}+\dot{h}^{2})z\delta f'-2(4+3h^{2}+\ddot{h}h+2\dot{h}^{2})\delta f \Big]=0
\end{align}
Inserting $\delta f(z)=\mu^{2\alpha} z^{2\alpha}$  in \eqref{EoM-g}, the equation of motion for $g(\theta)$ appears as,
\begin{align}
&h^{3}(1+h^{2}) \ddot{g}+4h^{2}\dot{h}(\alpha+2+h^{2})\dot{g}
+h\Big[2\alpha(2\alpha+1)(h^{2}+\dot{h}^{2})+(4\alpha+3+h^{2})(2\dot{h}^{2}-\ddot{h}h)
\nonumber\\
&-4(\alpha+1)(2\dot{h}^{2}+2h^{2}+3)-2h^{2}(2+h^{2})\Big]g
+[\alpha(h^{2}+\dot{h}^{2})-(4+3h^{2}+\ddot{h}h+2\dot{h}^{2})]=0.
\end{align}
Using the equation of motion for $h$ we can write
\begin{align}
&h^{3}(1+h^{2}) \ddot{g}+4h^{2}\dot{h}(1+h^{2})(\alpha+2+h^{2})\dot{g}
\nonumber\\
&+h\Big[2\alpha(2\alpha+1)(1+h^{2})(h^{2}+\dot{h}^{2})+(4\alpha+3+h^{2})(2+h^{2})(2\dot{h}^{2}+h^{2}+1)
\nonumber\\
&-4(\alpha+1)(1+h^{2})(2\dot{h}^{2}+2h^{2}+3)-2h^{2}(1+h^{2})(2+h^{2})\Big]g
\nonumber\\
&+\Big[\alpha(1+h^{2})(h^{2}+\dot{h}^{2})-2 \Big(2(1+h^{2})^{2}+\dot{h}^{2}h^{2}\Big)\Big]=0\,.
\end{align}

In order to find the universal term, we must extract the possible logarithmic or powering law divergences from the entanglement entropy. Suppose the deformation is small. By keeping the first order in the expansion of $f(z)$, and by substituting \eqref{delta-rho} in the entropy functional \eqref{SEE} we get, up to first order, the following perturbed entropy functional,
\begin{align}
&S_{EE}=\frac{L^{2}}{4G_{N}}\int_{\delta}^{z_{m}}dz \int_{\frac{-\Omega}{2}+\epsilon}^{\frac{\Omega}{2}-\epsilon}d\theta
\Big[\frac{\sqrt{\rho_{0}^{2}(1+\rho_0^{'2})+\dot{\rho_{0}}^{2}}}{z^{2}}
%\nonumber\\
-\frac{(\rho_{0}^{2}+\dot{\rho}^{2}) \delta f}{2z^{2}\sqrt{\rho_{0}^{2}(1+\rho_0^{'2})+\dot{\rho_{0}}^{2}}}
\nonumber\\
&\qquad +\frac{\rho_{0}^{2}\rho'_0\delta\rho'+\rho_{0}\delta\rho+\rho_0^{'2}\rho_{0}\delta\rho+
\dot{\rho_{0}}\dot{\delta\rho}}{z^{2}\sqrt{\rho_{0}^{2}(1+\rho_0^{'2})+\dot{\rho_{0}}^{2}}}\Big] \,.
\end{align}
Using integration by parts and the equation of motion for $\rho_{0}$, we find
\begin{align}\label{entropy-functional3}
&S_{EE}=\frac{L^{2}}{4G_{N}}\int_{\delta}^{z_{m}}dz \int_{\frac{-\Omega}{2}+\epsilon}^{\frac{\Omega}{2}-\epsilon}d\theta
\Bigg[\frac{\sqrt{\rho_{0}^{2}(1+\rho_0^{'2})+\dot{\rho_{0}}^{2}}}{z^{2}}
%\nonumber\\
-\frac{(\rho_{0}^{2}+\dot{\rho}^{2}) \delta f}{2z^{2}\sqrt{\rho_{0}^{2}(1+\rho_0^{'2})+\dot{\rho_{0}}^{2}}}
\nonumber\\
&\qquad +\partial_{\theta}\Big(
\frac{\dot{\rho_{0}}\delta\rho}{z^{2}\sqrt{\rho_{0}^{2}(1+\rho_0^{'2})+\dot{\rho_{0}}^{2}}}\Big)
+\partial_{z}\Big(
\frac{\rho_{0}^{2}\rho'_0\delta\rho}
{z^{2}\sqrt{\rho_{0}^{2}(1+\rho_0^{'2})+\dot{\rho_{0}}^{2}}}\Big)\Bigg]\,.
\end{align}
Now by substituting the ansatz $\rho(z,\theta)=\rho_{0}+\delta\rho=z/h(\theta)+z\delta fg(\theta)$ in \eqref{entropy-functional3} we reach to the following entropy functional,
\begin{equation}%\lable
S_{EE}=\frac{L^{2}}{2G_{N}}\Big (I_{1}+I_{2}+I_{3}+I_{4}\Big),
\end{equation}
where $I_{1}$, $I_{2}$, $I_{3}$, and $I_{4}$ are defined as:
\begin{align}
I_{1}&=\int_{\delta}^{z_{m}}\frac{dz}{z} \int_{0}^{\frac{\Omega}{2}-\epsilon}d\theta
\frac{\sqrt{1+h^{2}+\dot{h}^{2}}}{h^{2}}
\nonumber\\
 &=\int_{\delta}^{z_{m}}\frac{dz}{z} \int_{h_{0}}^{h_{c}}dh
\frac{\sqrt{1+h^{2}+\dot{h}^{2}}}{\dot{h}h^{2}} \,,\\
I_{2}&=\int_{\delta}^{z_{m}}dz\frac{\delta f}{2z} \int_{0}^{\frac{\Omega}{2}-\epsilon}d\theta
\frac{-(h^{2}+\dot{h}^{2}) }{h^{2}\sqrt{1+h^{2}+\dot{h}^{2}}}
\nonumber\\
&=\int_{\delta}^{z_{m}}dz\frac{\delta f}{2z} \int_{h_{0}}^{h_{c}}dh
\frac{-(h^{2}+\dot{h}^{2}) }{\dot{h}h^{2}\sqrt{1+h^{2}+\dot{h}^{2}}} \,, \\
I_{3}&=\int_{\delta}^{z_{m}}dz \int_{0}^{\frac{\Omega}{2}-\epsilon}d\theta
\partial_{\theta}\Big(
\frac{\dot{\rho_{0}}\delta\rho}{z^{2}\sqrt{\rho_{0}^{2}(1+\rho_0^{'2})+\dot{\rho_{0}}^{2}}}\Big)
\nonumber\\
&=\int_{\delta}^{z_{m}}dz\frac{\delta f}{z}\frac{-\dot{h}g(\theta)}{\sqrt{1+h^{2}+\dot{h}^{2}}}|_{\theta=\frac{\Omega}{2}-\epsilon} \,,\\
I_{4}&=\int_{\delta}^{z_{m}}dz \int_{0}^{\frac{\Omega}{2}-\epsilon}d\theta
\partial_{z}\Big(
\frac{\rho_{0}^{2}\rho'_0\delta\rho}
{z^{2}\sqrt{\rho_{0}^{2}(1+\rho_0^{'2})+\dot{\rho_{0}}^{2}}}\Big)
\nonumber\\
&=\int_{\delta}^{z_{m}}dz(\delta f') \int_{h_{0}}^{h_{c}}dh
\frac{g}{\dot{h}h\sqrt{1+h^{2}+\dot{h}^{2}}},
\end{align}
where we have changed the integration variable to $h$. We have also defined
$h_{0}=h(0)$, $h_{c}=h(\Omega/2-\epsilon)$, and used $\dot{h}_{0}(0)=0$ in getting boundary terms.

Now in order to single out the logarithmic divergences, we consider each term separately. By changing the derivative variable from $\theta$ to $h$ and by expressing $\ddot{g}(\theta)=d^{2}g/d\theta^{2}$ and $\dot{g}(\theta)=dg/d\theta$ in terms of $g''(h)=d^{2}g/dh^{2}$ and $g'(h)=dg/dh$, we can reach to the equation of motion for $g(h)$

\begin{align}%lable
&h^{3}(1+h^{2})^{2}(1+h^{2}-K^{2}h^{4})g''
\nonumber\\
&+h^{2}(1+h^{2})(2(1+h^{2})(2\alpha+3+2h^{2})-K^{2}h^{4}(4\alpha+8+5h^{2}))g'
\nonumber\\
&+h(2\alpha(2\alpha+1)((1+h^{2})^{2}-K^{2}h^{4})+(4\alpha+3+h^{2})(2+h^{2})(2+2h^{2}-K^{2}h^{4})
\nonumber\\
&-4(\alpha+1)(2(1+h^{2})^{2}+K^{2}h^{4})-2h^{2}(2+h^{2})K^{2}h^{4})g
\nonumber\\
&+[(1+h^{2})(\alpha+(\alpha-2)h^{2})-(\alpha+2)K^{2}h^{4}]=0
\end{align}
To extract the logarithmic divergence, we must consider the asymptotic behavior of $h$ and $g$. Depending on $\alpha$ we can choose the solutions for this equation such that $\rho$ becomes finite in the limit $h\rightarrow 0$ and $\delta\rightarrow 0$. Hence the asymptotic solution turns out to be
\begin{equation}\label{g}
  g=\frac{b}{h^{2\alpha+1}}+\frac{c_{1}}{h^{2\alpha-1}}+\frac{c_{3}}{h^{2\alpha-3}}+\frac{d_{1}}{h}+d_{2}h+\cdots
\end{equation}
where $b$, $c$'s and $d$'s are constants, and to avoid duplicated terms in (\ref{g}) we take $\alpha\neq1/2,1$,
\begin{align}
c_{1}&=\frac{1}{5}(2b+5b\alpha-2b\alpha^{2}) \nonumber\\ d_{1}&=\frac{-1}{2(3+2\alpha)} \nonumber\\
d_{2}&=\frac{2}{(3+2\alpha)(5+7\alpha+2\alpha^{2})}
\end{align}
and so on. Since $h_{c}=h(\Omega/2-\epsilon)$ and $z=\delta$, then UV cut-off expansion of $h_{c}$ becomes

\begin{equation}%\label{}
 h_{c}(\delta)=a_{1}\delta+a_{3}\delta^{3}+a_{5}\delta^{5}+e_{1}\delta^{2\alpha+1}+e_{2}\delta^{2\alpha+3}\cdots
\end{equation}
where $a_{i}$'s and $e_{j}$'s are coefficients that depend on $\alpha$,
\begin{align}
a_{1}&=\frac{1}{H}+b\mu^{2\alpha}{H^{2\alpha-1}} \nonumber\\ a_{3}&=c_{1}\mu^{2\alpha}{H^{2\alpha-3}} \nonumber\\
a_{5}&=c_{3}\mu^{2\alpha}{H^{2\alpha-5}}\nonumber\\ e_{1}&=\frac{d_{1}}{H} \mu^{2\alpha}\nonumber\\ e_{2}&=\frac{d_{2}}{H^{3}} \mu^{2\alpha}
\end{align}
and so on. In appendix $B$, we explicitly derive $g$ and $h_{c}$ for two values of $\alpha=1/2,1$.

Plugging this solution into the entropy functional is postponed to the next section where we add the Gauss-Bonnet gravity. Then the results of the Einstein gravity are included in the $\lambda_{GB}\rightarrow 0$ limit.

%---------------------------------------------------------------------------------------------------
\section{Gauss-Bonnet gravity}\label{sec4}

In the previous section, we studied the entanglement entropy for theories dual to the Einstein gravity. In this section we consider those which are dual to the Gauss-Bonnet gravity \cite{Ref43,Ref44}. We will consider the effect of the relevant perturbation on the entanglement entropy of these theories. After reviewing the holographic entanglement entropy for $CFT$s dual to Gauss-Bonnet gravity which was derived in \cite{Ref30,Ref45}, we proceed to the relevant perturbation of $CFT$s dual to the Gauss-Bonnet gravity in $AAdS$ backgrounds.

In $d=3$ dimensions, the bulk action for the Gauss-Bonnet gravity is written as
\begin{align}%\label{}
I=\frac{1}{16\pi G_{N}}\int d^{4}x\sqrt{g}\Big[\frac{6}{L^{2}}+R+\lambda_{GB}L^{2}\chi_{4}\Big],
\end{align}
where
\begin{align}%\label{}
\chi_{4}=R_{\mu\nu\rho\sigma}R^{\mu\nu\rho\sigma}-4R_{\mu\nu}R^{\mu\nu}+R^{2}
\end{align}
is the Euler density for four-dimensional manifolds. Hence the interaction term does not affect the gravitational equations of motion.

In this theory, by taking into account the effect of interaction term, the $RT$ prescription is improved as \cite{Ref30,Ref45}
\begin{equation}\label{SGB}
  S_{EE}= \frac{Area(m)}{4G_{N}}+\frac{L^{2}\lambda_{GB}}{2G_{N}}\int_{m}d^{2}x\sqrt{\gamma}\mathcal{R}
\end{equation}
where $m$ is the bulk minimal surface which is homologous to $V$ and $\partial_{m}$ matches the entangling surface $\partial_{V}$ on the the boundary. Also, $\gamma_{ij}$ and $\mathcal{R}$ are the induced metric and the intrinsic Ricci scalar of the bulk minimal surface, respectively.

Adding the interaction term leads to a topological contribution to the entropy functional with no effect on the universal term. Here $m$ is a two-dimensional sub-manifold and hence the second term in $RHS$ of \eqref{SGB} is proportional to the Euler characteristic which is a topological invariant of $m$, up to boundary terms, so it does not contribute to the equations of motion that determine the bulk minimal surface. We therefore use the solutions found in the previous section and substitute them in \eqref{SGB}.

% % % % % % % % % % % % % % % % % % % % % % % % % % %
\subsection{Pure $AdS$}
\label{sec41}

 We parameterize the bulk minimal surface as $\rho=\rho(z,\theta)$. Since the correction term in the \eqref{SGB} does not affect the profile of the bulk surface, we choose the bulk profile as before $\rho=z/h$. With this choice, the contribution of the correction term $\sqrt{\gamma}\mathcal{R}$ can be written as a total derivative
 \begin{equation}%\label{}
  \sqrt{\gamma}\mathcal{R}=\frac{d}{d\theta}\Big[\frac{2}{z}\frac{h\dot{h}}{\sqrt{1+h^{2}+\dot{h}^{2}}}\Big],
 \end{equation}
 then the holographic entanglement entropy becomes
 \begin{equation}\label{HEE}
   S_{EE}=\frac{L^{2}}{2G_{N}}\Big\{\int_{\delta}^{z_{m}}\frac{dz}{z}\int_{h_{0}}^{h_{c}} \frac{dh}{\dot{h}}\frac{\sqrt{1+h^{2}+\dot{h}^{2}}}{h^{2}}+2\lambda_{GB}\int_{\delta}^{z_{m}}\frac{dz}{z}\Big[\frac{\dot{h}}{h\sqrt{1+h^{2}+\dot{h}^{2}}}\Big]_{\theta=0}^{\frac{\Omega}{2}-\epsilon}\Big\}
 \end{equation}
in which $h_{0}=h(0)$, $h_{c}=h(\Omega/2-\epsilon)$, and $\dot{h}_{0}(0)=0$. Now we analyze the divergence of the above entropy functional. Near the boundary, the integrands  in the asymptotic limit, i.e. $h\rightarrow0$, behave as
\begin{align}%\label{}
  \frac{\sqrt{1+h^{2}+\dot{h}^{2}}}{\dot{h}h^{2}}&\thicksim-\frac{1}{h^{2}}-\frac{1}{2}K^{2}h^{2}+ \cdots \\
  \frac{\dot{h}}{h\sqrt{1+h^{2}+\dot{h}^{2}}}&\thicksim - \frac{1}{h}+\frac{1}{2}K^{2}h^{3}+ \cdots\,.
\end{align}
Now we can isolate the divergent part of the integrals in \eqref{HEE}. The first integral is the same as in \eqref{SEE0}, so we consider the second integral that is due to the effect of interaction term. Similar to procedure of section \ref{sec31}, we have
\begin{align}%lable
{I_{2}}=\int_{\delta}^{z_{m}}\frac{dz}{z}\frac{\dot{h}}{h\sqrt{1+h^{2}+\dot{h}^{2}}}\,,
\end{align}
 so we find
\begin{align}%\label{}
\frac{d{I_{2}}}{d\delta}=\frac{-1}{\delta}\frac{\dot{h}}{h\sqrt{1+h^{2}+\dot{h}^{2}}}=\frac{1}{\delta h_{c}}+\cdots=\frac{H}{\delta^{2}}+\cdots,
\end{align}
where in the last equality, we used $h_{c}(\delta)=\delta/H$. Hence, the final result for the holographic entanglement entropy becomes \cite{Ref27}
 \begin{equation}%\label{}
S_{EE}=\frac{L^{2}}{2G_{N}}\Big\{\frac{H}{\delta}(1-2\lambda_{GB})+\Big(-\int_{h_{0}}^{0}dh
(\frac{\sqrt{1+h^{2}+\dot{h}^{2}}}{\dot{h}h^{2}}+\frac{1}{h^{2}})+\frac{1}{h_{0}}\Big)\log(\delta)+\cdots\Big\}\,.
 \end{equation}
As we see, there is a logarithmic divergence only due to the singularity of the entangling surface and the effect of the interaction term is topological and modifies only the coefficient of the area law term\footnote{As was pointed out in \cite{Ref27}, see also appendix A, the $\lambda_{GB}$ dependence disappears by adding the boundary term. }.

% % % % % % % % % % % % % % % % % % % % % % % % % % % % % % % % %
\subsection{Asymptotic $AdS$}
\label{sec42}

In this subsection, we study the holographic entanglement entropy for a kink region in the asymptotic anti-de Sitter space time background in the Gauss-Bonnet gravity.

By equation \eqref{SGB}, the entanglement entropy is derived as
\begin{equation}\label{SEE-GB}
S_{EE}=\frac{1}{4G_{N}}\int_{\delta}^{z_{m}}dz \int_{\frac{-\Omega}{2}+\epsilon}^{\frac{\Omega}{2}-\epsilon}d\theta
\Big[\sqrt{\gamma}+2L^{2}\lambda_{GB}\sqrt{\gamma}\mathcal{R}\Big] = \frac{L^{2}}{2G_{N}}\int_{\delta}^{z_{m}}dz\int_{0}^{\frac{\Omega}{2}-\epsilon}d\theta \mathcal{L}
\end{equation}
where Lagrangian $\mathcal{L}$ defined as
\begin{align}
\mathcal{L}=&\frac{1}{z^{2}\sqrt{f}}\sqrt{(1+f\rho^{'2})\rho^{2}+\dot{\rho}^{2}}
+\frac{2\lambda_{GB}}{z^{2}\sqrt{f}}\frac{1}{\Big((1+f\rho^{'2})\rho^{2}+\dot{\rho}^{2}\Big)^{\frac{3}{2}}}
\nonumber\\
\times &\Big\{z f'[(\rho^{2}+\dot{\rho}^{2})^{2}-z\rho(\rho^{2}+2\dot{\rho}^{2}-\rho\ddot{\rho})\rho']-2z^{2}f\rho(\rho^{2}+2\dot{\rho}^{2}-\rho\ddot{\rho})\rho''
\nonumber\\
&-2f[\rho^{4}+\dot{\rho}^{4}-z\rho^{3}\rho'+z^{2}\dot{\rho}^{2}\rho^{'2}-2z\rho\rho'\dot{\rho}(\dot{\rho}+z\dot{\rho}')+\rho^{2}(2\dot{\rho}^{2}+z\ddot{\rho}\rho'+z^{2}\dot{\rho}'^{2})]
\nonumber\\
&+2f^{2}\rho^{2}\rho'[-\rho'(\rho^{2}+\dot{\rho}^{2}-z\rho\rho'+z\rho'\ddot{\rho}-2z\dot{\rho}\dot{\rho}')-z(\rho^{2}+\dot{\rho}^{2})\rho'']   \Big\}
\end{align}
with $\ddot{\rho}=\partial_{\theta}^{2}\rho$, $\rho''=\partial_{z}^{2}\rho$,  $\dot{\rho}'=\partial_{\theta}\partial_{z}\rho$, and $f(z)$ encodes the deviation from pure AdS background as in \eqref{AAdS}. As we see, the higher derivatives terms, like $\ddot{\rho}$, $\rho''$, and $\dot{\rho}'$ appear in the entropy functional. But due to the topological nature of the Gauss-Bonnet term, the equation of motion for $\rho(z,\theta)$ remains intact. As before, for small deformation we make the following ansatz,
\begin{equation}\label{ansatz}
  \rho(z,\theta)=\rho_{0}+\delta\rho=\frac{z}{h(\theta)}+\delta f g(\theta)z,\qquad  f=1+\delta f
\end{equation}
and by inserting it in the entropy functional \eqref{SEE-GB} we get, up to the leading order, the following perturbed entropy functional,
\begin{equation}%\label{}
S_{EE}=\frac{L^{2}}{2G_{N}}\int_{\delta}^{z_{m}}dz\int_{0}^{\frac{\Omega}{2}-\epsilon}d\theta\Big( \mathcal{L}_{0}(\rho_{0})+\delta \mathcal{L}\Big),
\end{equation}
where $\mathcal{L}_{0}(\rho_{0})$ comes from the unperturbed contribution of the original Lagrangian, and $\delta \mathcal{L}$ is due to the effect of the relevant perturbation defined as,
\begin{equation}%\label{}
\delta \mathcal{L}=\mathcal{L}_{f}+\mathcal{L}_{1}\delta\rho+\mathcal{L}_{2}\delta\rho'+\mathcal{L}_{3}\delta\dot{\rho}+\mathcal{L}_{4}\delta\ddot{\rho}+\mathcal{L}_{5}\delta\rho''+\mathcal{L}_{6}\delta\dot{\rho}'
\end{equation}
in which $\mathcal{L}_{f}$ is the contribution of $\delta f$, $\delta f'$, and independent of $g$ and its derivatives. The other terms come from the $\delta\rho$ and its various derivatives. Now we rewrite the derivative terms as
\begin{align}
\mathcal{L}_{2}\delta\rho'&=\partial_{z}(\delta\rho\mathcal{L}_{2})-\delta\rho\mathcal{L}_2',
\nonumber\\
\mathcal{L}_{3}\delta\dot{\rho}&=\partial_{\theta}(\delta\rho\mathcal{L}_{3})-\delta\rho\dot{\mathcal{L}_{3}},
\nonumber\\
\mathcal{L}_{4}\delta\ddot{\rho}&=\partial_{\theta}(\delta\dot{\rho}\mathcal{L}_{4})-\partial_{\theta}(\delta\rho\dot{\mathcal{L}_{4}})+\delta\rho\ddot{\mathcal{L}_{4}},
\nonumber\\
\mathcal{L}_{5}\delta\rho''&=\partial_{z}(\delta\rho'\mathcal{L}_{5})-\partial_{z}(\delta\rho\mathcal{L}_5')+\delta\rho\mathcal{L}_5'',
\nonumber\\
\mathcal{L}_{6}\delta\dot{\rho}'&=\partial_{z}(\delta\dot{\rho}\mathcal{L}_{6})-\partial_{\theta}(\delta\rho\mathcal{L}_6')+\delta\rho\dot{\mathcal{L}}_{6}',
\end{align}
where $\ddot{\mathcal{L}_{i}}=\partial_{\theta}^{2}\mathcal{L}_{i}$, $\mathcal{L}_i''=\partial_{z}^{2}\mathcal{L}_{i}$, and $\dot{\mathcal{L}}_{i}'=\partial_{\theta}\partial_{z}\mathcal{L}_{i}$.
Using these terms, the  entropy functional \eqref{SEE-GB} takes the form,
\begin{align}%\label{}
&S_{EE}=\frac{L^{2}}{2G_{N}}\int_{\delta}^{z_{m}}dz\int_{0}^{\frac{\Omega}{2}-\epsilon}d\theta\Big[ \mathcal{L}_{0}+\mathcal{L}_{f}+(\mathcal{L}_{1}-\mathcal{L}_2'-\dot{\mathcal{L}_{3}}+\ddot{\mathcal{L}_{4}}+\mathcal{L}_5''+\dot{\mathcal{L}}_{6}')\delta\rho
\nonumber\\
&\qquad+\partial_{z}(\delta\rho\mathcal{L}_{2}+\delta\rho'\mathcal{L}_{5}-\delta\rho\mathcal{L}_5'+\delta\dot{\rho}\mathcal{L}_{6})
+\partial_{\theta}(\delta\rho\mathcal{L}_{3}+\delta\dot{\rho}\mathcal{L}_{4}-\delta\rho\dot{\mathcal{L}_{4}}-\delta\rho\mathcal{L}_6')\Big]. \nonumber\\
\end{align}
In the above form of the entropy functional, the coefficient of $\delta\rho$ vanishes due to the equation of motion of $\rho_{0}(z,\theta)$. Then we have
\begin{align}\label{entropy-functional2}
&S_{EE}=\frac{L^{2}}{2G_{N}}\Big\{\int_{\delta}^{z_{m}}dz\int_{0}^{\frac{\Omega}{2}-\epsilon}d\theta\Big[ \mathcal{L}_{0}+\mathcal{L}_{f}+\partial_{z}(\delta\rho\mathcal{L}_{2}+\delta\rho'\mathcal{L}_{5}-\delta\rho\mathcal{L}_5'+\delta\dot{\rho}\mathcal{L}_{6})\Big]
\nonumber\\
&\qquad+\int_{\delta}^{z_{m}}dz(\delta\rho\mathcal{L}_{3}+\delta\dot{\rho}\mathcal{L}_{4}-\delta\rho\dot{\mathcal{L}_{4}}-\delta\rho\mathcal{L}_6')_{\theta=\frac{\Omega}{2}-\epsilon}\Big\}
\end{align}
Now by substituting the ansatz $\rho(z,\theta)=\rho_{0}+\delta\rho=z/h(\theta)+z\delta fg(\theta)$ in \eqref{entropy-functional2}, the entropy functional reduces to

\begin{align}%\label{}
&S_{EE}=\frac{L^{2}}{2G_{N}}\Big[\int_{\delta}^{z_{m}}dz\int_{h_{0}}^{h_{c}}\frac{dh}{\dot{h}}  \left(\mathcal{L}_{0}+\mathcal{L}_{f}+\mathcal{L}_{B}^{(z)}\right)+\int_{\delta}^{z_{m}}dz(\mathcal{L}_{B}^{(\theta)})_{\theta=\frac{\Omega}{2}-\epsilon}\Big].
\end{align}
where we have changed the integration variable to $h$. The various $\mathcal{L}_{i}$'s are defined as
\begin{align}
\mathcal{L}_{0}&=\frac{1}{z}(\frac{\sqrt{1+h^{2}+\dot{h}^{2}}}{h^{2}}+2\lambda_{GB}\frac{h(1+h^{2})\ddot{h}-(1+2h^{2}+\dot{h}^{2})\dot{h}^{2}}{h^{2}(1+h^{2}+\dot{h}^{2})^{\frac{3}{2}}})
\nonumber\\
&=\frac{1}{z}(\frac{\sqrt{1+h^{2}+\dot{h}^{2}}}{h^{2}}+2\lambda_{GB}\frac{d}{d\theta}\Big[\frac{\dot{h}}{h\sqrt{1+h^{2}+\dot{h}^{2}}}\Big])\equiv\frac{1}{z}(\hat{\mathcal{L}}^{(E)}_{0}(h)+\hat{\mathcal{L}}^{GB}_{0}(h)),
\nonumber\\
\mathcal{L}_{f}&=-\frac{\delta f}{2z}\Big[\frac{h^{2}+\dot{h}^{2} }{h^{2}\sqrt{1+h^{2}+\dot{h}^{2}}}+\frac{4\lambda_{GB}}{h^{2}(1+h^{2}+\dot{h}^{2})^{\frac{5}{2}}}\Big((2h^{4}+\dot{h}^{2}+\dot{h}^{4}+h^{2}(-1+3\dot{h^{2}})\Big)\dot{h}^{2}
\nonumber\\
&-(h^{2}+h^{4}+(3+h^{2})\dot{h}^{2})h\ddot{h}\Big)\Big]-\delta f'\frac{2\lambda_{GB}}{h^{2}(1+h^{2}+\dot{h}^{2})^{\frac{3}{2}}}\Big(h^{3}\ddot{h}-(\dot{h}^{2}+2h^{2})\dot{h}^{2}\Big)
\nonumber\\
&\equiv\frac{\delta f}{2z}\hat{\mathcal{L}}_{f}^{(1)}(h)+\delta f'\hat{\mathcal{L}}_{f}^{(2)}(h),
\nonumber\\
\mathcal{L}_{B}^{(z)}&=\delta f'\Big[\frac{g}{h\sqrt{1+h^{2}+\dot{h}^{2}}}-\frac{4\lambda_{GB}}{h(1+h^{2}+\dot{h}^{2})^{\frac{5}{2}}}\Big(2h\dot{h}\dot{g}(1+h^{2}+\dot{h}^{2})
\nonumber\\
&+g(h^{2}+2h^{4}+h^{6}+3\dot{h}^{2}+h^{2}\dot{h}^{2}+h^{4}\dot{h}^{2}+3\dot{h}^{4}+(h^{2}+h^{4}+(-3+h^{2})\dot{h}^{2})h\ddot{h})\Big)\Big]
\nonumber\\
&-z\delta f''\frac{4\lambda_{GB}}{h(1+h^{2}+\dot{h}^{2})^{\frac{3}{2}}}(h^{2}+h^{4}+\dot{h}^{2}+h^{3}\ddot{h})g
\nonumber\\
&\equiv\delta f'\hat{\mathcal{L}}_{1B}^{(z)}(h)+z\delta f''\hat{\mathcal{L}}_{2B}^{(z)}(h),
\nonumber\\
\mathcal{L}_{B}^{(\theta)}&=\frac{\delta f}{z}(-\frac{\dot{h}g(\theta)}{\sqrt{1+h^{2}+\dot{h}^{2}}}-4\lambda_{GB}\frac{h(1+h^{2})\dot{g}+\dot{h}(1-\dot{h}^{2})g}{(1+h^{2}+\dot{h}^{2})^{\frac{3}{2}}})\equiv\frac{\delta f}{z}\hat{\mathcal{L}}_{B}^{(\theta)}(h),
\end{align}
in which $\dot{h}=\partial_{\theta}h$, $\ddot{h}=\partial^{2}_{\theta}h$, and $\dot{g}=\partial_{\theta}g$. In the above expressions we used the equation of motion of $h$ \eqref{EoM-h}. Finally, the entropy functional reduces to
\begin{align}\label{SEE-GB2}
&S_{EE}=\frac{L^{2}}{2G_{N}}\Big\{\int_{\delta}^{z_{m}}\frac{dz}{z}\int_{h_{0}}^{h_{c}}\frac{dh}{\dot{h}} \hat{\mathcal{L}}_{0}(h)+\int_{\delta}^{z_{m}}dz\int_{h_{0}}^{h_{c}}\frac{dh}{\dot{h}}(\frac{\delta f}{2z}\hat{\mathcal{L}}_{f}^{(1)}(h)+\delta f'\hat{\mathcal{L}}_{f}^{(2)}(h))
\nonumber\\
&\qquad+\int_{\delta}^{z_{m}}dz\int_{h_{0}}^{h_{c}}\frac{dh}{\dot{h}}(\delta f'\hat{\mathcal{L}}_{1B}^{(z)}(h)+z\delta f''\hat{\mathcal{L}}_{2B}^{(z)}(h))+\int_{\delta}^{z_{m}}dz\frac{\delta f}{z}(\hat{\mathcal{L}}_{B}^{(\theta)}(h))_{\theta=\frac{\Omega}{2}-\epsilon}\Big\},
\end{align}
where $h_{0}=h(0)$ and $h_{c}=h(\Omega/2-\epsilon)$. So we have

\begin{equation}%\lable
S_{EE}=\frac{L^{2}}{2G_{N}}\Big (I_{0}+I_{f}+I^{(\theta)}_{B}+I^{(z)}_{B}\Big),
\end{equation}
in which $I_{0}$, $I_{f}$, $I^{(\theta)}_{B}$, and $I^{(z)}_{B}$ are defined as:
\begin{align}
I_{0}&=\int_{\delta}^{z_{m}}\frac{dz}{z}\int_{h_{0}}^{h_{c}}\frac{dh}{\dot{h}} \hat{\mathcal{L}}_{0}(h)
\nonumber\\
&=\int_{\delta}^{z_{m}}\frac{dz}{z}\int_{h_{0}}^{h_{c}}\frac{dh}{\dot{h}} \hat{\mathcal{L}}^{(E)}_{0}(h)+\int_{\delta}^{z_{m}}\frac{dz}{z}\hat{\mathcal{L}}^{(GB)}_{0}(h)
\nonumber\\
&=I^{(E)}_{0}+I^{(GB)}_{0}, \\
I_{f}&=\int_{\delta}^{z_{m}}dz\int_{h_{0}}^{h_{c}}\frac{dh}{\dot{h}}(\frac{\delta f}{2z}\hat{\mathcal{L}}_{f}^{(1)}(h)+\delta f'\hat{\mathcal{L}}_{f}^{(2)}(h))
\nonumber\\
&=\int_{\delta}^{z_{m}}dz\frac{\delta f}{2z}\int_{h_{0}}^{h_{c}}\frac{dh}{\dot{h}}\hat{\mathcal{L}}_{f}^{(1)}(h)+\int_{\delta}^{z_{m}}dz\delta f'\int_{h_{0}}^{h_{c}}\frac{dh}{\dot{h}}\hat{\mathcal{L}}_{f}^{(2)}(h)
\nonumber\\
&=I^{(1)}_{f}+I^{(2)}_{f}, \\
I^{(\theta)}_{B}&=\int_{\delta}^{z_{m}}dz\frac{\delta f}{z}(\hat{\mathcal{L}}_{B}^{(\theta)}(h))_{\theta=\frac{\Omega}{2}-\epsilon} \\
I^{(z)}_{B}&=\int_{\delta}^{z_{m}}dz\delta f'\int_{h_{0}}^{h_{c}}\frac{dh}{\dot{h}}\hat{\mathcal{L}}_{1B}^{(z)}(h)+\int_{\delta}^{z_{m}}dzz\delta f''\int_{h_{0}}^{h_{c}}\frac{dh}{\dot{h}}\hat{\mathcal{L}}_{2B}^{(z)}(h),
\nonumber\\
&=I^{(z)}_{1}+I^{(z)}_{2}\,.
\end{align}

Regarding the topological nature of $\sqrt{\gamma}\mathcal{R}$, the equations of motion of $h$ and $g$ remain intact. So their asymptotic behavior is unchanged, then the behavior of the integrands in \eqref{SEE-GB2} near the asymptotic boundary, $h\rightarrow0$, is
\begin{align}
\frac{\hat{\mathcal{L}}^{(E)}_{0}(h)}{\dot{h}}&\sim -\frac{1}{h^{2}}-\frac{1}{2}K^{2}h^{2}+\cdots
\nonumber\\
\hat{\mathcal{L}}^{(GB)}_{0}(h)&\sim 2\lambda_{GB}\left(-\frac{1}{h}+\frac{1}{2}K^{2}h^{3}+\cdots\right), \\
\frac{\hat{\mathcal{L}}_{f}^{(1)}(h)}{\dot{h}}&\sim \frac{1+4\lambda_{GB}}{h^{2}}+\frac{1}{2}(-1+36\lambda_{GB})K^{2}h^{2}+\cdots
\nonumber\\
\frac{\hat{\mathcal{L}}_{f}^{(2)}(h))}{\dot{h}}&\sim -\frac{2\lambda_{GB}}{h^{2}}+3\lambda_{GB}K^{2}h^{2}+\cdots,\\
\hat{\mathcal{L}}_{B}^{(\theta)}(h)&\sim (1-4\lambda_{GB})g+\frac{1}{2}(-1+20\lambda_{GB})K^{2}h^{4}g+\cdots,\\
\frac{\hat{\mathcal{L}}_{1B}^{(z)}(h)}{\dot{h}}&\sim (-1+36\lambda_{GB})K^{2}h^{3}g+\cdots
\nonumber\\
\frac{\hat{\mathcal{L}}_{2B}^{(z)}(h)}{\dot{h}}&\sim 4\lambda_{GB}K^{2}h^{3}g+\cdots \,.
\end{align}
Using these expressions, we isolate the divergent parts of integrands and make them finite. Hence, we reach to the following terms
\begin{align}\label{I0E}
I^{(E)}_{0}&=\int_{\delta}^{z_{m}}\frac{dz}{z}\int_{h_{0}}^{h_{c}}\frac{dh}{\dot{h}} \hat{\mathcal{L}}^{(E)}_{0}(h)
\nonumber\\
&=\int_{\delta}^{z_{m}}\frac{dz}{z}\int_{h_{0}}^{h_{c}}dh (\frac{\hat{\mathcal{L}}^{(E)}_{0}(h)}{\dot{h}}+\frac{1}{h^{2}})+\int_{\delta}^{z_{m}}\frac{dz}{z}(\frac{1}{h_{c}}-\frac{1}{h_{0}}), \\
I^{(GB)}_{0}&=\int_{\delta}^{z_{m}}\frac{dz}{z}\hat{\mathcal{L}}^{(GB)}_{0}(h),\\
I^{(1)}_{f}&=\int_{\delta}^{z_{m}}dz\frac{\delta f}{2z}\int_{h_{0}}^{h_{c}}dh(\frac{\hat{\mathcal{L}}_{f}^{(1)}(h)}{\dot{h}}- \frac{1+4\lambda_{GB}}{h^{2}})
-(1+4\lambda_{GB})\int_{\delta}^{z_{m}}dz\frac{\delta f}{2z}(\frac{1}{h_{c}}-\frac{1}{h_{0}}),\\
I^{(2)}_{f}&=\int_{\delta}^{z_{m}}dz\delta  f'\int_{h_{0}}^{h_{c}}dh(\frac{\hat{\mathcal{L}}_{f}^{(2)}(h)}{\dot{h}}+ \frac{2\lambda_{GB}}{h^{2}})
+2\lambda_{GB}\int_{\delta}^{z_{m}}dz\delta  f' (\frac{1}{h_{c}}-\frac{1}{h_{0}}),\\
I^{(\theta)}_{B}&=\int_{\delta}^{z_{m}}dz\frac{\delta f}{z}(\hat{\mathcal{L}}_{B}^{(\theta)}(h))_{\theta=\frac{\Omega}{2}-\epsilon}, \\
I^{(z)}_{B}&=\int_{\delta}^{z_{m}}dz\delta f'\int_{h_{0}}^{h_{c}}\frac{dh}{\dot{h}}\hat{\mathcal{L}}_{1B}^{(z)}(h)+\int_{\delta}^{z_{m}}dzz\delta f''\int_{h_{0}}^{h_{c}}\frac{dh}{\dot{h}}\hat{\mathcal{L}}_{2B}^{(z)}(h). \label{IBz}
\end{align}
Then we take a derivative with respect to $\delta$ and find the following expressions
\begin{align}%lable
\frac{dI^{(E)}_{0}}{d\delta}=&
\frac{1}{\delta}\Big[-\int_{h_{0}}^{0}dh
(\frac{\hat{\mathcal{L}}_{0}^{(E)}}{\dot{h}}+\frac{1}{h^{2}})+\frac{1}{h_{0}}\Big]+\frac{e_{1}}{a_{1}^{2}}\delta^{2\alpha-2}-\frac{1}{a_{1}\delta^{2}}+\cdots, \\
\frac{dI^{(GB)}_{0}}{d\delta}=&\frac{2\lambda_{GB}}{a_{1}\delta^{2}}-2\lambda_{GB}\frac{e_{1}}{a_{1}^{2}}\delta^{2\alpha-2}+\cdots\,,  \\
\frac{d{I^{(1)}_{f}}}{d\delta}=&\frac{(1+4\lambda_{GB})}{2a_{1}}\frac{\delta f}{\delta^{2}}-\frac{\delta f(\delta)}{2\delta}\Big(\int_{h_{0}}^{0}dh[\frac{\hat{\mathcal{L}}_{f}^{(1)} }{\dot{h}}-\frac{(1+4\lambda_{GB})}{h^{2}}]+\frac{(1+4\lambda_{GB})}{h_{0}}\Big)
\nonumber\\
&-(1+4\lambda_{GB})\frac{e_{1}}{2a_{1}^{2}}\delta f \delta^{2\alpha-2}-\frac{\delta f(-1+36\lambda_{GB})}{4}K^{2}a_{1}^{3}\delta^{2}+\cdots, \\
  \frac{d{{I^{(2)}_{f}}}}{d\delta}=&- \frac{2\lambda_{GB}}{a_{1}}\frac{\delta  f'}{\delta}+\cdots\,.
\end{align}
Using a similar procedure for the boundary terms we find that
\begin{align}%\label{}
\frac{dI^{(\theta)}_{B}}{d\delta}&=-\frac{\delta f}{\delta}\Big(1-4\lambda_{GB}+\frac{1}{2}(-1+20\lambda_{GB})K^{2}h^{4}\Big)g(\delta)+\cdots  \\
\frac{dI^{(z)}_{1B}}{d\delta}&=-\delta f' \int_{h_{0}}^{0}dh\frac{\hat{\mathcal{L}}_{1B}^{(z)}(h)}{\dot{h}}+\cdots  \\
\frac{dI^{(z)}_{2B}}{d\delta}&=-(\delta)\delta f'' \int_{h_{0}}^{0}dh\frac{\hat{\mathcal{L}}_{2B}^{(z)}(h)}{\dot{h}}+\cdots
\end{align}
so we have
\begin{align}\label{dItoddelta}
&\frac{d{I}}{d\delta}=\sum_i\frac{d{I_{i}}}{d\delta}
 \nonumber\\
&=\Big\{(1-2\lambda_{GB})\frac{-H}{\delta^{2}}+\Big(-\int_{h_{0}}^{0}dh
[\frac{\mathcal{L}_{0}^{(E)}}{\dot{h}}+\frac{1}{h^{2}}]+\frac{1}{h_{0}}\Big)\frac{1}{\delta}+\cdots\Big\} \nonumber\\\
&+\mu^{2\alpha}\Big\{2\lambda_{GB}\frac{bH^{2\alpha+1}}{\delta^{2}}+\frac{H}{2\delta^{2-2\alpha}}+2\lambda_{GB}\left(1+d_{1}-2\alpha \right)\frac{H}{\delta^{2-2\alpha}}-\frac{\left(1+4\lambda_{GB}(1-2\alpha)\right)}{2h_0\delta^{1-2\alpha}}
\nonumber\\
&-\Big(\frac{1}{2}\int_{h_{0}}^{0}dh[\frac{\hat{\mathcal{L}}_{f}^{(1)} }{\dot{h}}-\frac{(1+4\lambda_{GB})}{h^{2}}]
+2\alpha \int_{h_{0}}^{0}\frac{dh}{\dot{h}}
(\hat{\mathcal{L}}_{1B}^{z}+(2\alpha-1)\hat{\mathcal{L}}_{2B}^{z})\Big)\frac{1}{\delta^{1-2\alpha}}+\cdots\Big\}\,.
\end{align}
Up to now, we have identified the various kinds of divergences for a general situation. The integration of \eqref{dItoddelta} can be written as
\begin{equation}
S_{EE}=S_{EE}^{(0)}+\Delta S_{EE}^{(1)},
\end{equation}
where $ S_{EE}^{(0)} $ and $\Delta S_{EE}^{(1)} $ respectively represent the zeroth and leading order of the relevant perturbation. The former can be found as
\begin{align}
S_{EE}^{(0)} &=\frac{L^{2}}{2G_{N}}\Big[(1-2\lambda_{GB})\frac{H}{\delta}+\Big(-\int_{h_{0}}^{0}dh[\frac{\sqrt{1+h^{2}+\dot{h}^{2}}}{\dot{h}h^{2} }+\frac{1}{h^{2}}]+\frac{1}{h_{0}}\Big)\log(\delta)+\cdots\Big].
\end{align}
It exactly matches with results of \cite{Ref27}.
To find $\Delta S_{EE}^{(1)} $, it is needed to split various ranges of $\alpha $ as follows:\\

\textbf{i) $0<\alpha<\frac{1}{2}$}
\begin{align}\label{alpha-1}
\Delta S_{EE}^{(1)}=
&\frac{L^{2}\mu^{2\alpha}}{2G_{N}}\Big[-2\lambda_{GB}\Big(\frac{bH^{2\alpha+1}}{\delta}+\frac{1+d_{1}-2\alpha}{(1-2\alpha)}\frac{H}{\delta^{1-2\alpha}}\Big)-\frac{1}{2(1-2\alpha)}\frac{H}{\delta^{1-2\alpha}}+\cdots\Big],\nonumber\\
 \end{align}
this result shows that in the case of the Einstein gravity, $\lambda_{GB}=0$, singular term in the leading order is a power law singularity of order of $1/\delta^{1-2\alpha}$. In addition, there is a $1/\delta$ singularity in the Gauss-Bonnet gravity.

\textbf{ii) $\alpha=\frac{1}{2}$}
\begin{align}\label{alpha-2}
\Delta S_{EE}^{(1)}=
&\mu\frac{L^{2}}{2G_{N}}\Big[-2\lambda_{GB}\frac{bH^{2}}{\delta}+\frac{H}{4} (2-\lambda_{GB})\log(\delta)+\cdots\Big],
\end{align}
in this case, there is a logarithmic term both in the Einstein and the Gauss-Bonnet gravities. The $1/\delta$ singularity emerges in the Gauss-Bonnet case.

\textbf{iii) $\alpha >\frac{1}{2}$}
\begin{align}\label{alpha-3}
\Delta S_{EE}^{(1)}=
&\frac{L^{2}\mu^{2\alpha}}{2G_{N}}\Big[2\lambda_{GB}\frac{-bH^{2\alpha+1}}{\delta}+\cdots\Big].
\end{align}
Here when $\lambda_{GB}\rightarrow 0$, there is no contribution from the relevant perturbation to the entanglement entropy. In the Gauss-Bonnet gravity, a power law $1/\delta$ singularity contributes to the leading order of mass deformation $\mu$.

The summary of results are depicted in table \ref{table:questions} in which we introduce the angle dependent coefficient as
 \begin{align}%\label{}
   A(\Omega)&=\frac{1}{h_{0}}-\int_{h_{0}}^{0}dh
   \Big(\frac{\mathcal{L}_{0}^{(E)}}{\dot{h}}+\frac{1}{h^{2}}\Big)=\frac{1}{h_{0}}-\int_{h_{0}}^{0}dh\Big(\frac{\sqrt{1+h^{2}+\dot{h}^{2}}}{\dot{h}h^{2}}+\frac{1}{h^{2}}\Big)
   \nonumber\\
   &=\frac{1}{h_{0}}-\int_{h_{0}}^{0}dh\Big(\frac{h_{0}^{2}\sqrt{1+h^{2}}}{h^{2}\sqrt{(h_{0}^{2}-h^{2})(h_{0}^{2}+(1+h_{0}^{2})h^{2})}}+\frac{1}{h^{2}}\Big) \nonumber\\
   &\equiv\frac{2G_N}{L^2} a(\Omega),
 \end{align}
 where $a(\Omega)$ introduced in \eqref{SEE-aOmega}. Note that $A(\Omega)$ is the cut-off independent term that only appears due to the singularity of the entangling surface \cite{Ref25}. The dependence of $A(\Omega)$ on the opening angle is through the dependence of $h_{0}$ on $\Omega$, such that we can write \cite{Ref27}
 \begin{equation}\label{Omega}
 \Omega=\int_{-\frac{\Omega}{2}}^{\frac{\Omega}{2}}d\theta=\int_{0}^{h_{0}}dh
 \frac{2h^{2}\sqrt{1+h_{0}^{2}}}{\sqrt{1+h^{2}}\sqrt{(h_{0}^{2}-h^{2})(h_{0}^{2}+(1+h_{0}^{2})h^{2})}}.
 \end{equation}

\begin{table}[ht]
	\caption{Summary of results for $\frac{2G_N}{L^2}S_{EE}$. Note that contributions of the Gauss-Bonnet and the relevant perturbation are additive to the Einstein gravity. }\label{table:questions}
	\begin{center}
		\begin{tabular}{|l l| l| l|}
			\hline
			& & Einstein Gravity & Gauss-Bonnet contribution \\
			\hline
			$\mu=0$ & & $\frac{H}{\delta}-A(\Omega)\log\frac{H}{\delta}$ & $-2\lambda_{GB}\frac{H}{\delta}$ \\
			$\mu\neq 0$ & $0<\alpha<\frac{1}{2}$ & $\frac{-1}{2(1-2\alpha)}\frac{\mu^{2\alpha}H}{\delta^{1-2\alpha}}$ & $2\lambda_{GB}\mu^{2\alpha}H^{2\alpha}\left(-\frac{1+d_{1}-2\alpha}{1-2\alpha}\frac{H^{1-2\alpha}}{\delta^{1-2\alpha}}-\frac{bH}{\delta}\right) $\\
			& $\alpha=\frac{1}{2}$ & $-\frac{\mu H}{2}\log\frac{H}{\delta}$ & $2\lambda_{GB}\mu H\left(\frac{-bH}{\delta}+\frac{1}{8}\log\frac{H}{\delta}\right) $ \\
			& $\alpha>\frac{1}{2}$ & 0 & $2\lambda_{GB}\mu^{2\alpha}H^{2\alpha}\left(\frac{-bH}{\delta}\right) $ \\
			\hline
		\end{tabular}
	\end{center}
\end{table}

\section{The renormalized entanglement entropy}
\label{sec5}

As was mentioned in the introduction, the entanglement entropy is a $UV$ divergent quantity, such that the leading term scales with the area of the entangling surface and sub-leading terms exhibit the power law divergences. The coefficients of those terms are scheme dependent, but there is certain sub-leading term which is logarithmic or constant, depending on dimension of space time, so that its coefficient is universal and describes the character of the underlying quantum field theory.

In order to have a well-defined quantity we must somehow get rid of the $UV$ divergences and extract a finite and universal quantity, which is physically meaningful. In other words, the entanglement entropy must be renormalized. One of the procedures to dealing with this problem is the renormalized entanglement entropy ($REE$) which was introduced in \cite{Ref20} and is based on the differentiation of the entanglement entropy with respect to the characteristic length of the entangling region\footnote{See also \cite{Ref46} which is based on the area renormalization of the entangling surfaces.}. In the spirit of \cite{Ref20}, the renormalized entanglement entropy is derived by applying the differential operator ($d=3$), $D(R)=R\partial_{R}-1$, on the entanglement entropy $S(R)$:
\begin{equation}
 \mathcal{F}(R)=R\partial_{R}S(R)-S(R),
 \end{equation}
where $R$ is the characteristic size of the entangling surface. The $\mathcal{F}(R)$ quantity is finite in a renormalizable $3d$ quantum field theory which has a well-defined $UV$ fixed point. This procedure is defined for a scalable surface like a sphere that can be scaled without any shape deformation. In this sense our singular surface is a scalable one. Let us firstly consider the Einstein gravity and by analogy define the renormalized entanglement entropy as
\begin{align}\label{ren-1}
\mathcal{F}(H)&=H\partial_{H}(H\partial_{H}-1)S(H)
\end{align}
where $H$ is the characteristic scale of the entangling surface. In order to regularize divergences as well as possible finite terms, we compute explicitly the entanglement entropy from the relations \eqref{I0E}-\eqref{IBz}. At the $UV$ fixed point the entanglement entropy has the form :
 \begin{equation}%\label{}
S_{EE}=\frac{L^{2}}{2G_{N}}\Big\{\frac{H}{\delta}-A(\Omega)\log(\frac{H}{\delta})+\cdots\Big\}\,.
 \end{equation}
 So the renormalized entanglement entropy derived as
 \begin{equation}\label{renorm-F}
  \mathcal{F}_{UV}=\frac{L^{2}}{2G_{N}}A(\Omega)=a(\Omega).
  \end{equation}
 As we see this term is finite, positive, independent of the $UV$ divergence, and so has a well-defined continuum limit. It is $H$-independent and specified by the universal part of the entanglement entropy, so it is physically meaningful. Under the relevant deformation the entanglement entropy takes the following form

 \begin{align}\label{alpha-11}
   S_{EE}=\frac{L^{2}}{2G_{N}}\Big\{\frac{H}{\delta}-A(\Omega)\log(\frac{H}{\delta})+
 M_{E}(h_{0},\alpha)\mu^{2\alpha}H^{2\alpha}
  &-\frac{1}{2(1-2\alpha)}\frac{\mu^{2\alpha}H}{\delta^{1-2\alpha}}\Big\}\,\nonumber\\
   \end{align}
where $h_{0}=h(0)$ which depends on the angle $\Omega$ through (\ref{Omega}) and $\alpha=3-\Delta$ with $\Delta$ the scaling dimension of the relevant operator.  $M_{E}(h_{0},\alpha)$ is defined as the sum of finite terms that come from the relations \eqref{I0E}-\eqref{IBz}:
\begin{align}
M_{E}(h_{0},\alpha)=M_{0}^{E}(h_{0},\alpha)+M_{f}(h_{0},\alpha)+M_{B}^{\theta}(h_{0},\alpha)
+M_{B}^{z}(h_{0},\alpha).
\end{align}
These terms are finite and functions of $h_{0}$ and $\alpha$, which with some manipulations we can derive
\begin{align}
M_{E}(h_{0},\alpha)=&-\frac{h_{0}^{2\alpha}}{4\alpha} \int_{h_{0}}^{0}dh[
\frac{(h^{2}+\dot{h}^{2}) }{\dot{h}h^{2}\sqrt{1+h^{2}+\dot{h}^{2}}}+\frac{1}{h^{2}}]+\frac{h_{0}^{2\alpha-1}}{4\alpha(1-2\alpha)}\nonumber\\
&+h_{0}^{2\alpha} \int_{h_{0}}^{0}dh[
\frac{g }{\dot{h}h\sqrt{1+h^{2}+\dot{h}^{2}}}].
\end{align}
  From (\ref{ren-1}) the renormalized entanglement entropy can be derived as
   \begin{align}\label{F-Einstein}
       \mathcal{F}(H)=\mathcal{F}_{UV}-\mathcal{M}_{E}(h_{0},\alpha)\mu^{2\alpha}H^{2\alpha}.
     \end{align}
where the second term denotes the leading correction due to the relevant perturbation and 
\begin{equation}
\mathcal{M}_{E}(h_{0},\alpha)=\frac{L^{2}}{2G_{N}}2\alpha (1-2\alpha)M_{E}(h_{0},\alpha)\,.
\end{equation} 
As we see $\mathcal{F}(H)$ reduces to that of the $UV$ fixed point in the limit
\begin{equation}
\mathcal{F}(H)\rightarrow  \mathcal{F}_{UV}, \qquad H\rightarrow 0.
\end{equation}
In the special case $\alpha=1/2$, or equivalently $\Delta=5/2$, $M_{E}(h_{0},\alpha)$ is divergent while $\mathcal{M}_{E}(h_{0},\alpha)$ is finite and we have
	\begin{align}%\label{}
    \mathcal{F}(H)=\mathcal{F}_{UV}-\frac{\mu H}{2}.
    \end{align}
In conclusion, we can interpret $\mathcal{F}(H)$ as the measure of the entanglement at the scale $H$.
Note, in the above renormalization approach/scheme we have used the minimal prescription. Indeed, one can show that there are several renormalization operators with the same renormalized entanglement entropy at the $UV$ fixed point but different RG flows. For example consider the following operators
\begin{align}
\frac{(-1)^{n-1}}{(n-1)!}H^n\partial^n_H (H\partial_H-1),
\end{align}
where there are infinitely many linear combinations of these operators with different $n$ that reproduce the same $\mathcal{F}_{UV}$ while they give different flows away the $CFT$ fixed point. This ambiguity in introducing the renormalization operator was reported earlier in \cite{Ref47,Ref48}. To choose the correct operator, one needs to apply some extra physical conditions. Besides, let us first consider the most general operator by changing the variable to $u=\log(H/\delta)$ then $H\partial_H=\partial_u$. The renormalization operator in (\ref{ren-1}) is a second order one made of $\partial_u$. So in general we can write it as a polynomial,\footnote{In this and the following equations by dots we mean an arbitrary finite number of terms.}
\begin{align}\label{polyD}
D(\partial_u)=\alpha_0+\alpha_1\partial_u+\alpha_2\partial_u^2+\cdots,
\end{align}
with $\alpha_i$'s constant. Now look at the bare entropy in table \ref{table:questions}. The singularities appear as $e^u$ and $u$ where the last one is the universal log term and to regularize them one need to consider,  respectively, $1$ and zero roots for the polynomial operator(\ref{polyD}). Thus we have    
\begin{align}
D(\partial_u)=\mathcal{P}(\partial_u)(\partial_u-1)\partial_u,
\end{align}
where 
%the numerical factor is written such that to produce the correct renormalized entropy at the $UV$ fixed point and
$\mathcal{P}(\partial_u)$ is any polynomial in the following form,
\begin{align}
\mathcal{P}(\partial_u)=(1+\beta_1\partial_u+\beta_2\partial_u^2+\cdots).
\end{align}
Now, the ambiguity in introducing the renormalization operator is the freedom in choosing the coefficients of $\mathcal{P}(\partial_u)$. To fix it we need to consider the renormalization flow. Let us take the criteria that $\mathcal{F}$ satisfies the $F$-theorem in the fashion of \cite{Ref17,Ref18}. Then it should be decreasing away from the $UV$ fixed point. It implies that $\mathcal{M}(h_0,\alpha)$ in either (\ref{F-Einstein}) or (\ref{F-GB}) should be positive. It follows that
\begin{align}
\mathcal{M}(h_0,\alpha)&=\frac{L^{2}}{2G_{N}}(2\alpha)(1-2\alpha)\mathcal{P}(2\alpha)M(h_0,\alpha)>0 
\end{align} 
then $\mathcal{P}(2\alpha)$ should be chosen such that $\mathcal{M}(h_0,\alpha)$ to be positive and finite term.

Based on the above results, we have a family of operators and it seems there is no more physical constraint to choose the proper renormalization operator. We see, in the vicinity of a $UV$ fixed point, these are the most restrictive criteria which can be imposed on $D(\partial_u)$. To achieve a unique renormalization operator, we propose that the extra condition may be to consider the renormalization flow downward to an $IR$ fixed point. In this way it seems one can determine which polynomial $\tilde{\mathcal{P}}(\partial_u)$ correctly flows between the $UV$ and the $IR$ fixed points\footnote{We assume there is an $IR$ fixed point.}. This needs the $IR$ calculations which we postpone to future works.

The above computations are related to the Einstein gravity. Similarly, for the Gauss-Bonnet gravity we can write
\begin{equation}\label{S1}
S_{EE}=\frac{L^{2}}{2G_{N}}\Big\{(1-2\lambda_{GB})\frac{H}{\delta}-A(\Omega)\log(\frac{H}{\delta})+\cdots\Big\}\,.
 \end{equation}
So using (\ref{ren-1}), the renormalized entanglement entropy  becomes
\begin{equation}
  \mathcal{F}_{UV}=\frac{L^{2}}{2G_{N}}A(\Omega)=a(\Omega).
  \end{equation}
As we see this term is the same as \eqref{renorm-F} with a well-defined continuum limit.  Under the relevant deformation the entanglement entropy takes the following form
  \begin{align}\label{alpha-12}
  \Delta S_{EE}=&\frac{L^{2}}{2G_{N}}M_{GB}(h_{0},\alpha)\mu^{2\alpha}H^{2\alpha}
  \nonumber\\
\qquad&+\frac{L^{2}\mu^{2\alpha}}{2G_{N}}\Big[-2\lambda_{GB}\Big(\frac{bH^{2\alpha+1}}{\delta}+\frac{1+d_{1}-2\alpha}{(1-2\alpha)}\frac{H}{\delta^{1-2\alpha}}\Big)-\frac{1}{2(1-2\alpha)}\frac{H}{\delta^{1-2\alpha}}+\cdots\Big],\nonumber\\
   \end{align}
similar  to the previous discussion,  $M_{E}(h_{0},\alpha)$ is defined as the sum of finite terms coming from the relations \eqref{I0E}-\eqref{IBz}.
\begin{align}
M_{GB}(h_{0},\alpha)=M_{0}^{E}(h_{0},\alpha)+M_{0}^{GB}(h_{0},\alpha)+M_{f}(h_{0},\alpha)+M_{B}^{\theta}(h_{0},\alpha)
+M_{B}^{z}(h_{0},\alpha).
\end{align}
But, in this case we must refine the differential operator of renormalization in order to remove the effect of $bH^{2\alpha+1}/{\delta}$ term. So, we define
\begin{align}\label{ren-2}
\mathcal{F}(H)&= \frac{-1}{(2\alpha+1)}H\partial_{H}(H\partial_{H}-1)(H\partial_{H}-(2\alpha+1))S(H),
\end{align}
as a renormalized entanglement entropy and find
 \begin{align}\label{F-GB}
       \mathcal{F}(H)=\mathcal{F}_{UV}-\mathcal{M}_{GB}(h_{0},\alpha)\mu^{2\alpha}H^{2\alpha}.
     \end{align}
where the second term denotes the leading correction due to the relevant perturbation where 
\begin{equation}
\mathcal{M}_{GB}(h_{0},\alpha)=\frac{L^{2}}{2G_{N}}\frac{2\alpha (1-2\alpha)}{(2\alpha+1)}M_{GB}(h_{0},\alpha)\,.
\end{equation} 
Again in the small $H$ limit, the $\mathcal{F}(H)$ reduces to that of the $UV$ fixed point
 \begin{equation}
  \mathcal{F}(H)\rightarrow  \mathcal{F}_{UV}, \qquad H\rightarrow 0.
 \end{equation}
In the special case $\alpha=1/2$ we find that
 \begin{align}%\label{}
  \mathcal{F}(H)=\mathcal{F}_{UV}-\frac{\mu H}{8}(2-\lambda_{GB}).
 \end{align}
Notice that both operators in (\ref{ren-1}) and (\ref{ren-2}) give the same $\mathcal{F}_{UV}$. Again similar to the Einstein gravity, there is a family of renormalization operators that doing the same thing which similar to the previous discussion we can write    
\begin{align}
D(\partial_u)=\frac{-1}{(2\alpha+1)}\mathcal{P}(\partial_u)(\partial_u-1)(\partial_u-(2\alpha+1))\partial_u,
\end{align}
where the numerical factor is written such that to produce the correct renormalized entropy at the $UV$ fixed point and $\mathcal{P}(\partial_u)$ is any polynomial in the following form,
\begin{align}
\mathcal{P}(\partial_u)=(1+\beta_1\partial_u+\beta_2\partial_u^2+\cdots).
\end{align}
Again decreasing $\mathcal{F}$ away from the $UV$ fixed point implies that 
\begin{align}
\mathcal{M}(h_0,\alpha)&=\frac{L^{2}}{2G_{N}}\frac{(2\alpha)(2\alpha-1)}{(2\alpha+1)}\mathcal{P}(2\alpha)M(h_0,\alpha)>0 
\end{align} 

In this case, more investigations are needed in dual higher derivatives theories to better understand their flow.

%%%%%%%%%%%%%%%%%%%%%%%%%%%%%%%%%%%%%%%%%%%%%%%%%%%%%%%%%%%%%%%%%%%%%%%%%%%%%%%%%%
\section{Discussion}
\noindent In this paper we studied a $CFT$ which is perturbed by a relevant operator and identified various divergence structures that may appear in the holographic entanglement entropy of a kink region.

There are two kinds of data contributions in the calculation of $EE$, the geometric data and field theoretic data, and each contribution may be separately specified. As we have seen in the previous sections, in three dimensions, in the absence of relevant perturbation there are two kinds of divergences; the power law of order $1/\delta$ and the logarithmic divergences. The later is due to the singularity of the entangling surface and reflects the geometric contribution to the entanglement entropy, and the angle dependent coefficient $a(\Omega)$ of this term, which satisfies some properties \cite{Ref23,Ref24,Ref25}, encodes some features of the underlying $CFT$ \cite{Ref26,Ref27,Ref28}.

In the relevant perturbation case, there are also two kinds of divergences due to the effect of relevant perturbation which are of field theoretic type. These divergences appeared for some special values of the scaling dimension of the relevant operator $\Delta$. At the first order of perturbation, only for the special value of $\Delta=5/2(=(d+2)/2)$ we had a logarithmic divergence, whereas for other values of $\Delta$ we had power law divergence of the order $1/\delta^{1-2\alpha}$. The appearance of the logarithmic divergence at $\Delta=(d+2)/2$ is independent of the singularity of the entangling surface and was firstly reported in \cite{Ref35}. This indicates that the relevant perturbation and the surface singularity have two distinguished log contribution to the entanglement entropy.

In order to consider the effect of the relevant perturbation in higher derivative gravities, we chose the Gauss-Bonnet gravity which is topological in four-dimensional space time. As expected there is no effect on the logarithmic term and only the coefficient of the area law term is modified. But, after adding boundary term, all effects of the Gauss-Bonnet term disappear and the entanglement entropy remains unaffected. However, in the relevant perturbation, even though the Gauss-Bonnet term is topological, two kinds of $\lambda_{GB}$ dependent terms appear. As shown in the appendix $A$, these terms are unaffected by the surface term.

In summary, in the Einstein gravity, there are two separate contributions due to the singularity and the relevant perturbation. The corner contribution appears only in the Einstein gravity and the pure $CFT$, i.e. it is independent of $\mu$ and $\lambda_{GB}$. Contributions of the Gauss-Bonnet and relevant perturbation are additive to the Einstein gravity. On the other hand, there is a log contribution due to the combination of the Gauss-Bonnet gravity and the relevant perturbation. This effect is absent in the pure $CFT$ and happens again at $\Delta=5/2$.

The contribution of the logarithmic term is universal in the sense that its value is independent of the precise details of the $UV$ regulator, so the appearance of that kind of terms helps us to probe the characteristics of the underlying theory.

We also introduced the renormalized entanglement entropy for the kink region in the vacuum. It is positive, finite, universal, and well-defined in continuum limit. It is intrinsic to the underlying conformal field theory at the $UV$ fixed point and is an $H$ independent constant $\mathcal{F}_{UV}$. In the relevant perturbation, it is sensitive to the scale of the size $H$ of the entangled region and $\mathcal{F}(H)\rightarrow  \mathcal{F}_{UV}$ as $ H\rightarrow 0$. But, as we seen, we faced by a family of the renormalization operators that all of them yield the same $\mathcal{F}_{UV}$ at the $CFT$ fixed point. Unfortunately, based on our results it seems that there is not enough physical reason to choose the unique differential operator and it needs to be more investigated in order to define an appropriate and unique renormalized entanglement entropy away from fixed points. We speculate that the renormalization group flow downward to an $IR$ fixed point, if any, may distinguish a unique renormalization operator. In this way seems one can determine which polynomial $\tilde{\mathcal{P}}(\partial_u)$ correctly flows between the $UV$ and the $IR$ fixed points.

At the end it is worth to mention few comments. Firstly, as was shown in \cite{Ref35}, given the form of the expansion of $f(z)$ around $z=0$ in the source deformation, $f(z)=1+\sum_{m=2}^{\infty}c_{m}z^{m\alpha}$, and noticing the form of the expansion of the minimal surface, in order to have a logarithmic term  we must have $\alpha=(d-2)/m$ and it immediately follows that the conformal dimension of the dual operator must be $\Delta=d-(d-2)/m$. So the appearance of the logarithmic term depends on the order of perturbation expansion of the metric and the bulk minimal surface, and it only arises for $\alpha\leq d/2-1$ or equivalently, the scaling dimension of the relevant operator corresponds to $\Delta\geq d/2+1$. So as we see we may have a set of $\Delta$'s that leads to a logarithmic term depending on the order of the perturbation. Hence in the smooth case, there is a family of operators with the scaling dimension $\Delta=d-(d-2)/m$ which produce the corresponding logarithmic term in the entanglement entropy. So when one perturbs the $CFT$ with any of the listed above operators, the logarithmic term shows up itself in some order of perturbation. Therefore, it seems that in our case too, if we perturb the $CFT$ with any of single operators in the above family, then we will have the logarithmic term in some order of expansion. But, note that it is natural to expect that in a given conformal field theory, the perturbation is done by few operators not all of them in the above family. So few logarithmic terms in some order of perturbation may appear. 

Secondly, although the operators with scaling dimension  $\Delta=d-(d-2)/m$ with $m\geq2$ is special and does not arise in a free $CFT$ or a scale invariant field theory, but in the interacting theories, the existence of such operators are possible. Since the scaling dimension receives quantum corrections through the anomalous dimension, it might be possible to have scalar operators with scaling dimension $\Delta=d-(d-2)/m$ in an interacting $CFT$ \cite{Ref40}.

Finally, it would be interesting to investigate these divergence structures in higher dimensions, as smooth case, and for various kinds of singularities \cite{Ref49}.

%--------------------------------------------------------------------------
\vspace{1cm}
\noindent {\large {\bf Acknowledgment} }  MG would like to thank Sepideh Mohammadi for encouragement and valuable comments. Authors would like to thank Ali Imaanpur for reading the manuscript and the referee of JHEP for useful comments and motivating to add section 5.

%---------------------------------------------------------------------------
\appendix
\section{Appendix}
In this appendix, we will show that the logarithmic term due to the relevant perturbation of HEE of Gauss-Bonnet is unaffected by the addition of a surface term.
As was denoted in \cite{Ref27}, one can add the boundary term to the entropy functional \eqref{SGB},
\begin{equation}%\label{}
S_{B}= \frac{L^{2}\lambda_{GB}}{G_{N}}\int_{\partial m} dy \sqrt{\tilde{\gamma}}\mathcal{K},
\end{equation}
where $\partial m$ denotes the boundary of Ryu-Takayanagi at the cut-off surface $z=\delta$, and $\tilde{\gamma}$ and $\mathcal{K}$ are the determinant of the induced metric and the trace of extrinsic curvature, respectively.

To consider this boundary, it is convenient to change the independent coordinate system, and write the $\rho$ as one of coordinates induced on the bulk minimal surface. So we have $z=z(\rho,\theta)$. The induced metric on the boundary of the bulk minimal surface parametrized as $\{\rho=\rho, \theta=\Omega/2, z=\delta\}$, is written as

\begin{equation}%\label{}
 ds ^{2}= \frac{L^{2}}{z^{2}}d\rho^{2}\;,
\end{equation}
so we find that
\begin{equation}%\label{}
 \sqrt{\tilde{\gamma}}\sim \frac{L}{\delta}\;.
\end{equation}
Moreover, the associated normal vector and the trace of the extrinsic curvature can be written as
\begin{equation}%\label{}
n\sim(0,-\frac{z}{L}\sqrt{f(z)},0,0),
\end{equation}
and
\begin{equation}%\label{}
\mathcal{K}= \frac{\sqrt{f}}{L}\sim\frac{1}{L}(1-\frac{1}{2}\mu^{2\alpha}\delta^{2\alpha}\big)\,.
\end{equation}
Finally, the contribution of the boundary term is written as

\begin{align}
\int_{\frac{\delta}{h_{0}}+\cdots}^{H}d\rho\mathcal{K}  \sqrt{\tilde{\gamma}}\sim\frac{H}{\delta}-\frac{1}{2}H\mu^{2\alpha}\delta^{2\alpha-1}+\cdots
 \end{align}
where dots in the lower limit in the integral denote the higher order terms of $\delta$ and in the right hand side stand for the regular terms. Substituting the above relation in the surface term, we reach to
\begin{equation}%\label{}
S_{B}= \frac{L^{2}}{2G_{N}}2\lambda_{GB}\frac{H}{\delta} \,.
\end{equation}
As we see the effect of the boundary term is only on the area term, and does not affect the universal term.

%------------------------------------------------------------------------
\section{Appendix}
In this appendix, we derive the explicit form of $g$, $h_{c}$ and the holographic entanglement entropy for two values of $\alpha$ :

$\mathbf{i)\; \Delta=\frac{5}{2},\;\; (\alpha=\frac{1}{2})}$: \\
First we consider $\Delta=5/2$, in this range $\alpha=1/2$. So we find
\begin{equation}%\label{}
  g=\frac{b}{h^{2}}-\frac{1}{8h}+\frac{4b}{5}+\frac{1}{18}h+\cdots
\end{equation}
and
\begin{equation}%\label{}
 h_{c}(\delta)=\frac{1}{H}(1+\mu bH)\delta-\frac{\mu}{8H}\delta^{2}+\frac{4b\mu}{5H^{2}}\delta^{3}+\cdots
\end{equation}
Note that we have kept only the leading order in $\mu$ at any order in $\delta$. So the divergence structures that appears are
\begin{align}%lable
\frac{dI_{1}^{(1)}}{d\delta}&=
-\frac{1}{\delta}\int_{h_{0}}^{0}dh
[\frac{\sqrt{1+h^{2}+\dot{h}^{2}}}{\dot{h}h^{2}}+\frac{1}{h^{2}}]+\frac{1}{2H^{3}}K^{2}(1+3\mu bH)\delta^{2}+\cdots,\\
\frac{dI_{1}^{(2)}}{d\delta}&=-\frac{H}{\delta^{2}}+\frac{1}{h_{0}}\frac{1}{\delta}+\mu(\frac{bH^{2}}{\delta^{2}}-\frac{H}{8}\frac{1}{\delta})  \\
\frac{dI_{2}^{(1)}}{d\delta}&=\frac{\mu}{2}\int_{h_{0}}^{0}dh[\frac{(h^{2}+\dot{h}^{2}) }{\dot{h}h^{2}\sqrt{1+h^{2}+\dot{h}^{2}}}+\frac{1}{h^{2}}]+\frac{\mu}{4H^{3}}K^{2} \delta^{3}+\cdots, \\
\frac{dI_{2}^{(2)}}{d\delta}&= \frac{H\mu}{2}\frac{1}{\delta}-\frac{1}{h_{0}}\frac{\mu}{2}+\cdots
\end{align}
And for the boundary terms we have:
\begin{align}%\label{}
\frac{d{I_{3}}}{d\delta}&=-\mu\Big(bH^{2}\frac{1}{\delta^{2}}-\frac{H}{8}\frac{1}{\delta}+\frac{4b}{5}\Big)+\cdots \\
\frac{d{I_{4}}}{d\delta}&= -\mu\int_{h_{0}}^{0}dh
\frac{g}{\dot{h}h\sqrt{1+h^{2}+\dot{h}^{2}}}+\cdots
\end{align}
Again we have kept only the leading order in $\mu$. So we find HEE for AAdS geometry,
\begin{align}
 S_{EE}=&\frac{L^{2}}{2G_{N}}\Big[\frac{H}{\delta}+\Big(-\int_{h_{0}}^{0}dh[\frac{\sqrt{1+h^{2}+\dot{h}^{2}}}{\dot{h}h^{2} }+\frac{1}{h^{2}}]+\frac{1}{h_{0}}\Big)\log(\delta)+\cdots\Big]
 \nonumber\\
 &+\mu\frac{L^{2}}{2G_{N}}\Big[\frac{H}{2} \log(\delta)+\cdots\Big]
 \nonumber\\
 \end{align}
  As we expect the first term denotes the unperturbed part of EE, but second term is due to the relevant perturbation of the boundary theory in the  geometry of bulk minimal surface.

$\mathbf{ii)\; \Delta=1,\;\; (\alpha=1)}$: \\
Now we consider $\Delta=1$, in this range $\alpha=1$. So we find
\begin{equation}%\label{}
  g=\frac{b}{h^{3}}+\frac{10b-1}{10h}+\frac{1+5bK^{2}}{35}h+\cdots
\end{equation}
and
\begin{equation}%\label{}
 h_{c}(\delta)=\frac{1}{H}(1+\mu^{2}H^{2}b)\delta+\frac{(10b-1)\mu^{2}}{10H}\delta^{3}+\frac{(1+5bk^{2})\mu^{2}}{35H^{3}}\delta^{5}+\cdots
\end{equation}
Note that here the leading correction is second order in $\mu$ at any order in $\delta$. So the divergence structures that appears are
\begin{align}%lable
&\frac{dI_{1}^{(1)}}{d\delta}=
-\frac{1}{\delta}\int_{h_{0}}^{0}dh
[\frac{\sqrt{1+h^{2}+\dot{h}^{2}}}{\dot{h}h^{2}}+\frac{1}{h^{2}}]+\frac{1}{2H^{3}}K^{2}(1+3bH^{2}\mu^{2})\delta^{2}+\cdots,
\end{align}
\begin{equation}%\label{}
\frac{dI_{1}^{(2)}}{d\delta}=-\frac{H}{\delta^{2}}+\frac{1}{h_{0}}\frac{1}{\delta}+\mu^{2}\frac{bH^{3}}{\delta^{2}}
\end{equation}
\begin{align}%\label{}
\frac{dI_{2}^{(1)}}{d\delta}=\frac{\mu^{2}}{2}\Big(\int_{h_{0}}^{0}dh[\frac{(h^{2}+\dot{h}^{2}) }{\dot{h}h^{2}\sqrt{1+h^{2}+\dot{h}^{2}}}+\frac{1}{h^{2}}]\Big)\delta+\frac{\mu^{2}}{4H^{3}}K^{2} \delta^{4}+\cdots,
\end{align}
\begin{equation}%\label{}
\frac{dI_{2}^{(2)}}{d\delta}= \frac{H\mu^{2}}{2}-\frac{1}{h_{0}}\frac{\mu^{2}}{2}\delta+\cdots
\end{equation}
And for the boundary terms we have:
\begin{align}%\label{}
\frac{d{I_{3}}}{d\delta}=-\mu^{2}\Big(bH^{3}\frac{1}{\delta^{2}}+\frac{10b-1}{10}H+\frac{1+5bK^{2}}{35H}\delta\Big)+\cdots
\end{align}
\begin{align}%\label{}
&\frac{d{I_{4}}}{d\delta}= -2\mu^{2}\Big(\int_{h_{0}}^{0}dh
\frac{g}{\dot{h}h\sqrt{1+h^{2}+\dot{h}^{2}}}\Big)\delta+\cdots
\end{align}
Then we find
\begin{align}
 S_{EE}=&\frac{L^{2}}{2G_{N}}\Big[\frac{H}{\delta}+\Big(-\int_{h_{0}}^{0}dh[\frac{\sqrt{1+h^{2}+\dot{h}^{2}}}{\dot{h}h^{2} }+\frac{1}{h^{2}}]+\frac{1}{h_{0}}\Big)\log(\delta)+\cdots\Big]
 \end{align}
 As we expect, in this case, only the unperturbed part of EE appears and the effect of the relevant perturbation of the boundary theory disappears.
%--------------------------------------------------------------------------

\end{document}